\documentclass[aps,twocolumn,english,notitlepage]{revtex4-2}

\usepackage[T1]{fontenc}
\usepackage[latin9]{inputenc}
\usepackage{babel}
\usepackage{enumitem}
\usepackage{pifont}
\usepackage{verbatim}
\usepackage{float}
\usepackage{amsmath}
\usepackage{amsthm}
\usepackage{amssymb}
\usepackage{graphicx}
\usepackage{appendix}
\usepackage[dvipsnames]{xcolor}
\usepackage{array}
\usepackage{multirow}
\usepackage{ulem}
\usepackage{hyperref}
\hypersetup{
	colorlinks=true,
	linkcolor=blue,
	filecolor=magenta,      
	urlcolor=cyan,
	citecolor=red
}
\newcolumntype{P}[1]{>{\centering\arraybackslash}p{#1}}

\usepackage{subfigure,bbm}

\usepackage{mathtools}

\DeclarePairedDelimiter\floor{\lfloor}{\rfloor}

\makeatother

\newcommand{\be}{\begin{equation}}
\newcommand{\ee}{\end{equation}}
\newcommand{\ba}{\begin{align}}
\newcommand{\ea}{\end{align}}

\newcommand{\sign}{\text{sign}}

\makeatletter

\providecommand{\tabularnewline}{\\}

\begin{document}

\title{Coherent manipulation of graph states composed of finite-energy\\ Gottesman-Kitaev-Preskill-encoded qubits}

\author{Kaushik P. Seshadreesan$^{1,2}$}
\email{kausesh@pitt.edu}
\author{Prajit Dhara$^{1,3}$}
\author{Ashlesha Patil$^{1}$}
\author{Liang Jiang$^{4}$}
\author{Saikat Guha$^{1}$}

\affiliation{$^{1}$Wyant College of Optical Sciences, The University of Arizona, Tucson,
AZ 85721, USA}
\affiliation{$^{2}$Department of Informatics and Networked Systems, School of Computing and Information, University of Pittsburgh, Pittsburgh, PA 15260, USA}
\affiliation{$^{3}$Birla Institute of Technology \& Science Pilani, Pilani, Rajasthan 333031, India}
\affiliation{$^{4}$Pritzker School of Molecular Engineering, The University of Chicago, Chicago, IL 60637, USA}

\begin{abstract}
    Graph states are a central resource in measurement-based quantum information processing. In the photonic qubit architecture based on Gottesman-Kitaev-Preskill (GKP) encoding, the generation of high-fidelity graph states composed of realistic, finite-energy approximate GKP-encoded qubits thus constitutes a key task. 
    We consider the finite-energy approximation of GKP qubit states given by a coherent superposition of shifted finite-squeezed vacuum states, where the displacements are Gaussian distributed. 
    We present an exact description of graph states composed of such approximate GKP qubits as a coherent superposition of a Gaussian ensemble of randomly displaced ideal GKP-qubit graph states. 
    Using standard Gaussian dynamics, we track the transformation of the covariance matrix and the mean displacement vector elements of the Gaussian distribution of the ensemble under tools such as GKP-Steane error correction and fusion operations that can be used to grow large, high-fidelity GKP-qubit graph states. 
    The covariance matrix elements capture the noise in the graph state due to the finite-energy approximation of GKP qubits, while the mean displacements relate to the possible absolute shift errors on the individual qubits arising conditionally from the homodyne measurements that are a part of these tools. 
    Our work thus pins down an exact coherent error model for graph states generated from truly finite-energy GKP qubits, which can shed light on their error correction properties.
\end{abstract}

%\date{\today}

\maketitle

\section{Introduction}
Photonic quantum technologies~\cite{SP19,FSS18,DS15,OFV09} provide a promising avenue for realizing quantum information processing in practice. 
A number of scalable architectures~\cite{AYFM18,TF17,Menicucci14,Rudolf17} for fault-tolerant universal quantum computation using realistically imperfect, noisy photonic elements, are being actively pursued experimentally~\cite{Xanadu20,Asavanant19,CMP14}. 
Photonic, noisy intermediate-scale quantum (NISQ) processors are being utilized for demonstrations of quantum advantage over classical computations, e.g., in the boson sampling problem~\cite{Zhong2020-bb,Tillmann13}. 
Moreover, photonics is ubiquitously used in quantum communications~\cite{GT07} and quantum sensing~\cite{PBGW18} since photons form the most natural choice for carriers of quantum information.

Quantum information is most commonly encoded in the photonic domain in the discrete, finite degrees of freedom of single photons such as their polarization or propagation paths, or transverse spatial modes~\cite{DLBPA11,EFKZ18}, or frequency~\cite{CSGF20,LL17,RDJFT14} or temporal modes~\cite{ASSVML17,BRSR15}, or time bins~\cite{SMAKKZT19,JPKHSW14}. 
Deterministic generation of single photons forms the key challenge in realizing these encodings. 
Alternatively, encodings in the continuous, infinite quadrature degrees of freedom of spatial, frequency and temporal modes of the bosonic field have also been considered~\cite{WPGCRSL12}. 
The class of Gaussian continuous variable (CV) states, such as the coherent states and squeezed states are easily generated using lasers and quantum nonlinear optics. 
CV quantum states are especially suited for the paradigm of measurement-based quantum computation since highly entangled CV multimode Gaussian graph states~\cite{Asavanant19,YYKSSMF16,CMP14} can be generated very efficiently. 
However, implementing universal quantum logic~\cite{MR11,RGMMG03}, or any useful non-trivial quantum information processing task such as entanglement distillation~\cite{ESP02} or quantum error correction~\cite{NFC09} over CV states requires non-Gaussian elements such as photon number resolving detection or third or higher order optical nonlinearities~\cite{LB99}.

In 2001, Gottesman, Kitaev and Preskill~\cite{GKP01}, introduced a hybrid encoding of quantum information in a bosonic mode, where an error-corrected qubit (more generally a qudit) is encoded in the continuous quadrature degrees of freedom of a bosonic mode. 
The GKP qubit, is protected against continuous, small displacement errors, which is critical to realize fault-tolerant quantum computation using CV quantum states and measurements. 
The resilience of GKP qubits against small displacement errors also makes them resistant to photon loss errors that are encountered in quantum communications. In fact, among all possible finite-dimensional subspace-encodings over the CV, infinite dimensional Hilbert space of a bosonic mode~\cite{AND18}, the GKP qubit encoding is close to the optimal encoding for quantum capacity of Gaussian thermal loss channels with average photon number constraint~\cite{Noh2019-ne}. This makes them suitable for error correction-based quantum repeaters~\cite{Fukui2020-hx,Rozpedek2020-zs}. 
GKP qubit states are sufficiently non-Gaussian that all qubit-level Clifford operations can be deterministically and efficiently implemented using linear optics and coherent homodyne detection~\cite{MvLGWRn06}. 
The key challenge in working with GKP qubits is that since they are ideal, unnormalized states, they can only be approximately realized in practice. 
There have been a few different proposals to realize approximate GKP qubit states and  experimental implementations as well in recent works~\cite{Campagne-Ibarcq2020,De_Neeve2020-fm,Fluhmann2019-eh,MBGM17,TW16}. 

Given a supply of approximate GKP qubit states, the generation of CV GKP graph states has important applications in measurement-based quantum computation as well as in all-optical quantum repeaters, where the graph states play the role of quantum memories~\cite{Fukui2020-hx,PKES17}. 
This task was investigated in Ref.~\cite{FTOF18}. 
However, the approximate GKP qubits were modeled as incoherent mixtures of ideal GKP qubits shifted by Gaussian distributed random displacements~\cite{Wang2019} that are strictly speaking still infinite energy states and hence unphysical. 
On the other hand, a coherent superposition of shifted finite-squeezed vacuum states, where the displacements are Gaussian distributed is a truly finite-energy approximation of a GKP qubit. 
An exact description of graph states based on such finite-energy, approximate GKP qubit pure states has been missing. This is accomplished in the present work. 
We start by considering the error wavefunction description of such finite-energy approximate GKP qubit pure states given by a coherent superposition of a Gaussian ensemble of ideal GKP-qubit states that are randomly displaced in phase space. 
Based on this, we represent a finite-energy GKP qubit graph state as a coherent superposition of a Gaussian ensemble of ideal GKP qubit graph states that are randomly displaced in phase space, characterized by the mean displacement vector and the covariance matrix of the Gaussian distribution for the random displacements. 
Using standard tools from Gaussian dynamics~\cite{GLS16}, we track the transformation of these characteristics under the GKP-Steane error correction protocol and graph fusion operations that are used to conditionally prepare large, high-fidelity graph states composed of individual GKP qubit pure states. 
An important merit of the description is that it provides a coherent error model for the GKP qubit graph states, which can be used to study their best error correction properties.
A coherent error model could, e.g., be useful in designing all-optical GKP-encoding-based quantum repeaters based on graph states, and more generally for quantum computing with realistic, finite-energy, GKP qubit encoding.
The paper is organized as follows.  In Sec.~\ref{prelim}, we briefly review the GKP encoding of a qubit in a bosonic field mode along with its finite-energy approximations. 
In Sec.~\ref{GKPgraph}, we describe finite-energy GKP graph states. 
In Sec.~\ref{singlequbit QEC}, we discuss an error correction procedure due to Steane~\cite{Steane1997-op}, which is widely used for GKP qubits. 
In Sec.~\ref{fusion_ops}, we discuss two fusion operations, that are used to merge two subgraphs or modify parts of a graph. 
In Sec.~\ref{discussions}, we apply our description of finite-energy GKP qubit graph states to a graph state generation protocol to grow graph states. We illustrate how the description provides an accurate acccount of the noise and errors that build up in the graph state in the protocol.

\section{Gottesman-Kitaev-Preskill (GKP) qubits}
\label{prelim}
Consider a bosonic mode described by its annihilation and creation operators $\hat{a}$ and $\hat{a}^\dagger$, such that $[\hat{a},\hat{a}^\dagger]=1.$ 
The corresponding Hermitian quadrature operators are given by $\hat{q}=(\hat{a}+\hat{a}^\dagger)/\sqrt{2},$ 
$\hat{p}=(\hat{a}-\hat{a}^\dagger)/(\sqrt{2}i)$, where we have chosen $\hbar=1$. 
The eigenstates of these operators are Fourier related as $|q\rangle=\frac{1}{\sqrt{2\pi}}\int dp \exp(ipq)|p\rangle$ and $|p\rangle=\frac{1}{\sqrt{2\pi}}\int dp \exp(-iqp)|q\rangle.$ 
The symmetrically-ordered displacement operator for the mode is defined as
\begin{align}
\hat{D}\left(\frac{u+iv}{\sqrt{2}}\right)&=e^{-iu \hat{p}+iv\hat{q}};\ u,v\in\mathbb{R}\label{displacement operator}\\
&=e^{i\frac{uv}{2}}\hat{X}(u)\hat{Z}(v)=e^{-i\frac{uv}{2}}\hat{Z}(v)\hat{X}(u)\label{XZ_ZX_displ}\\
\textrm{where\ } \hat{Z}(v)&=e^{iv\hat{q}},\ 
\hat{X}(u)=e^{-iu\hat{p}}.
\end{align}
The operators $\hat{X}(2\sqrt{\pi})$, $\hat{Z}(2\sqrt{\pi})$ commute, i.e., they can be simultaneously diagonalized. 
The ideal GKP qubit is defined as the 2-D subspace stabilized by these operators~\cite{GKP01}. 
The logical bit-flip and phase flip operators for this qubit are defined as $\overline{X}=\hat{X}(\sqrt{\pi}), \ \overline{Z}=\hat{Z}(\sqrt{\pi})$. 
The ideal eigenstates of these operators form the bases for a GKP qubit and are given by
\begin{align}
|\overline{\sign((-1)^{k_p})}\rangle&=\sum_{n=-\infty}^{+\infty}|(2n+k_p)\sqrt{\pi}\rangle_p,\ k_p\in\{0,1\}\label{p-GKP_Ideal}\\
|\overline{k_q}\rangle&=\sum_{n=-\infty}^{+\infty}|(2n+k_q)\sqrt{\pi}\rangle_q,\ k_q\in\{0,1\}\label{q-GKP_Ideal},
\end{align}
respectively. It can be easily shown using the Poisson summation formula that the states of (\ref{q-GKP_Ideal}) are uniform coherent superpositions of the states in (\ref{p-GKP_Ideal}) with appropriate phases, and vice versa. 
Moreover, the Wigner function of a uniform incoherent mixture of the bases states in (\ref{p-GKP_Ideal}) or (\ref{q-GKP_Ideal}), i.e., e.g., $\left(|\overline{0}\rangle\langle\overline{0}|+|\overline{1}\rangle\langle\overline{1}|\right)/2$, is a collection of delta function peaks that lie on a square lattice of spacing $\sqrt{\pi}$ in phase space. For this reason, the qubit that they define is referred to as a square-lattice GKP qubit.

Since the ideal square-lattice GKP qubit-basis states defined above are infinite superpositions of periodically displaced, infinite energy quadrature eigenstates, they are unnormalizable and unphysical. 
We can define finite-energy approximations of GKP qubit states as superpositions of periodically displaced, finitely squeezed vacuum states of variance $\sigma^2/2$ (where $\sigma^2=1$ corresponds to the vacuum), weighted by a Gaussian envelope function of variance $2/\sigma^2$, which therefore have finite energy, e.g.,
\begin{align}
|\widetilde{k_q}\rangle&\propto\sum_{n=-\infty}^{+\infty}e^{-\frac{\sigma^2((2n+k_q)\sqrt{\pi})^2}{2}}\hat{X}((2n+k_q)\sqrt{\pi})\nonumber\\
&\times\frac{1}{(\pi\sigma^2)^{1/4}}\int_{-\infty}^{+\infty} dq\, e^{-\frac{q^2}{2\sigma^2}}|0\rangle_q,\ k_q\in\{0,1\},\label{finite_sq_01_states}
\end{align}
and likewise the $|\widetilde{k_p}\rangle$ states. 
When $\sigma\ll\sqrt{\pi}$, the normalization constants of the above states are $\approx \sqrt{2\sigma}$.

\begin{figure}
    \centering
    \includegraphics[width=0.45\textwidth]{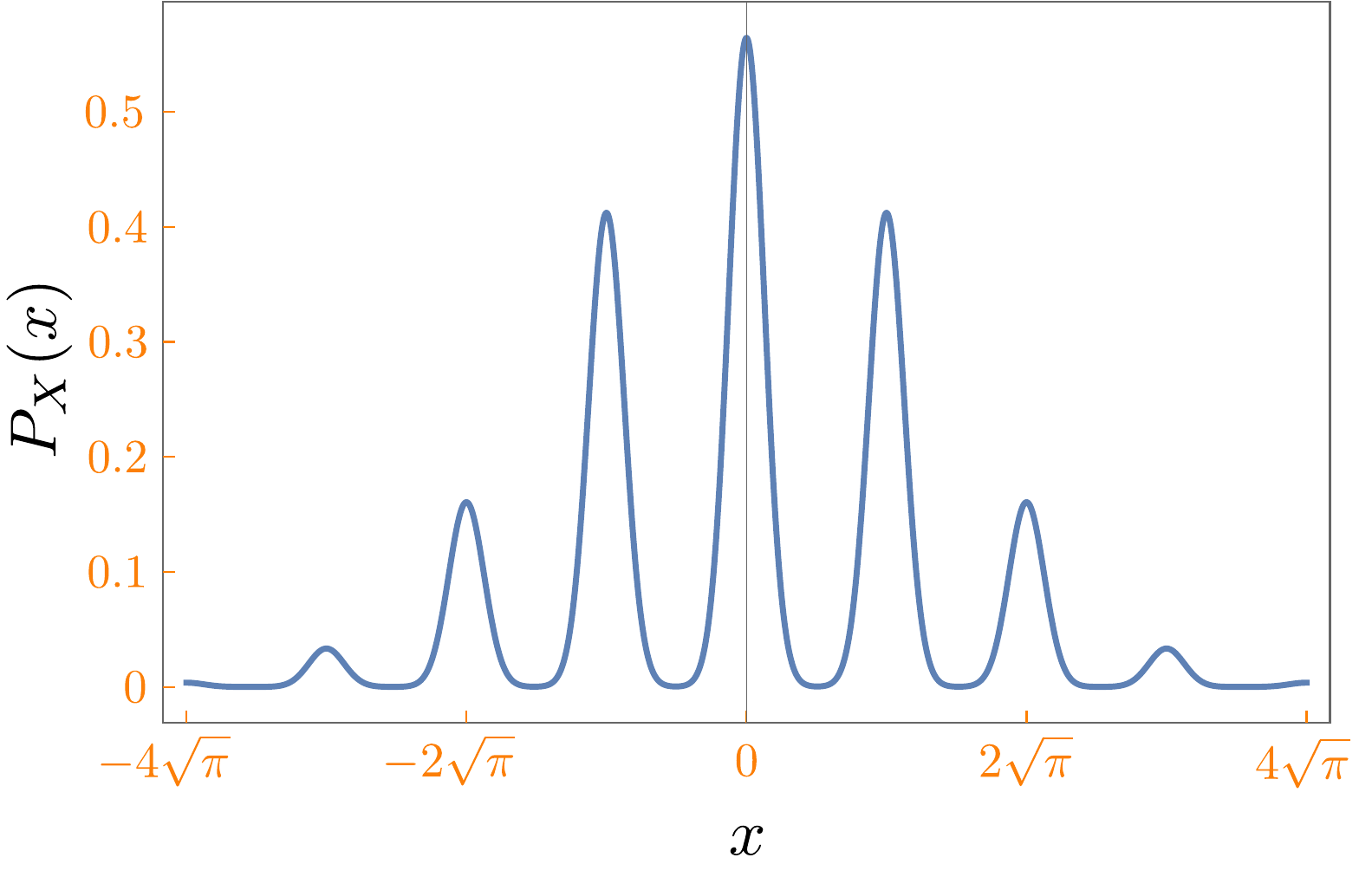}
    \caption{Probability distribution of outcomes from $q-$quadrature measurement of a finite-energy GKP qubit $(|\widetilde{0}\rangle+|\widetilde{1}\rangle)/\sqrt{2}$ for a squeezing value of 10 dB.}
    \label{fig:qmeasdistri}
\end{figure}

When such a finitely squeezed approximate GKP qubit state, say the state $|\widetilde{0}\rangle$ is measured along the $q-$quadrature using homodyne detection, the probability distribution of the outcomes and its approximation when $\sigma\ll\sqrt{\pi}$ are given by
\begin{align}
P_{X}(x)&=|\langle x|\tilde{0}\rangle|^2\approx\bigg|\sum_{n=-\infty}^{+\infty}\sqrt{2\sigma}\,e^{-\frac{\sigma^2(2n\sqrt{\pi})^2}{2}}\frac{e^{-\frac{(x-2n\sqrt{\pi})^2}{2\sigma^2} }}{(\pi\sigma^2)^{1/4}} \bigg|^2\nonumber\\
&\approx\sum_{n=-\infty}^{+\infty}(2\sigma)e^{-4\pi\sigma^2n^2} \frac{e^{-\frac{(x-2n\sqrt{\pi})^2}{\sigma^2}}}{\sqrt{\pi\sigma^2}}\\%\ \textrm{when\ } \sigma<<\sqrt{\pi}\\
&=\sum_{n=-\infty}^{+\infty}P_N[n]\, P_Q(x-2\sqrt{\pi}n),
\end{align}
where
\begin{align}
P_N[n]&=(2\sigma)e^{-4\pi\sigma^2n^2},\ n\in\mathbb{Z}\label{PN_homodyne}\\
%\Rightarrow P_N[2\sqrt{\pi}n]&=\sqrt{\frac{\sigma^2}{\pi}} e^{-\frac{n^2}{1/\ssigma}},\ n\in\mathbb{Z}\label{PN_homodyne}\\
P_Q(q)&=\frac{e^{-\frac{q^2}{\sigma^2}}}{\sqrt{\pi\sigma^2}},\ q\in\mathbb{R}\label{PQ_homodyne}
\end{align}
are Gaussian distributions of an integer-valued random variable $N$ and a real-valued random variable $Q$, respectively. We denote these random variables and their distributions by the following shorthand notation that highlights the distribution (along with the field), and the mean and variance: $N\sim\mathcal{G}_\mathbb{Z}(0,1/(8\pi\sigma^2))$ and $Q\sim\mathcal{G}_\mathbb{R}(0,\sigma^2/2)$, where $\sigma^2\in\mathbb{R}$. Rescaling the random variable $N$ by $2\sqrt{\pi}$, we have a real-valued random variable $2\sqrt{\pi}N\sim\mathcal{G}_\mathbb{R}(0,1/(2\sigma^2))$. Thus, the outcome $X$ is a random variable given by
\begin{align}
X=2\sqrt{\pi} N+Q,
\end{align}
whose distribution $P_{X}(x), \ x\in\mathbb{R}$ is given by the convolution of the distributions of $2\sqrt{\pi}N$ and $Q$. 
Likewise, when a coherent superposition state of the form $(|\widetilde{0}\rangle+|\widetilde{1}\rangle)/\sqrt{2}$ is measured with $q-$homodyne detection, we have
\begin{align}
&P_{X}(x)=\frac{1}{2}|\langle x|\tilde{0}\rangle+\langle x|\tilde{1}\rangle|^2\nonumber\\
&\approx\frac{1}{2}\bigg|\sum_{n=-\infty}^{+\infty}\sqrt{2\sigma}e^{-\frac{\sigma^2(2n\sqrt{\pi})^2}{2}}\frac{e^{-\frac{(x-2n\sqrt{\pi})^2}{2\sigma^2} }}{(\pi\sigma^2)^{1/4}}\nonumber\\
&+\sum_{n'=-\infty}^{+\infty}\sqrt{2\sigma}e^{-\frac{\sigma^2((2n'+1)\sqrt{\pi})^2}{2}}\frac{e^{-\frac{(x-(2n'+1)\sqrt{\pi})^2}{2\sigma^2} }}{(\pi\sigma^2)^{1/4}}\bigg|^2\\
&=\frac{1}{2}\bigg|\sum_{n=-\infty}^{+\infty}\sqrt{2\sigma}e^{-\frac{\sigma^2(n\sqrt{\pi})^2}{2}}\frac{e^{-\frac{(x-n\sqrt{\pi})^2}{2\sigma^2} }}{(\pi\sigma^2)^{1/4}}\bigg|^2\\
&\approx\sum_{n=-\infty}^{+\infty}\sigma e^{-\pi\sigma^2n^2} \frac{e^{-\frac{(x-n\sqrt{\pi})^2}{\sigma^2}}}{\sqrt{\pi\sigma^2}}\\%\ \textrm{when\ } \sigma<<\sqrt{\pi}\\
&=\sum_{n=-\infty}^{+\infty}P_N[n]\, P_Q(x-\sqrt{\pi}n),
\end{align}
where
\begin{align}
P_N[n]&=\sigma e^{-\pi\sigma^2n^2},\ n\in\mathbb{Z}\label{PN_homodyne_conjugate_meas}\\
P_Q(q)&=\frac{e^{-\frac{q^2}{\sigma^2}}}{\sqrt{\pi\sigma^2}},\ q\in\mathbb{R}\label{PQ_homodyne_conjugate_meas}
\end{align}
are Gaussian distributions of an integer-valued random variable $N\sim\mathcal{G}_\mathbb{Z}(0,1/(2\pi\sigma^2))$ and a real-valued random variable $Q\sim\mathcal{G}_\mathbb{R}(0,\sigma^2/2)$, respectively. 
That is, the outcome is a random variable
\begin{align}
X=\sqrt{\pi} N+Q,
\end{align}
whose distribution $P_{X}(x), \ x\in\mathbb{R}$ is the convolution of distributions of $\sqrt{\pi}N\sim\mathcal{G}_\mathbb{R}(0,1/(2\sigma^2))$ and $Q\sim\mathcal{G}_\mathbb{R}(0,\sigma^2/2)$. 
Figure~\ref{fig:qmeasdistri} plots this outcome distribution for the case where the constituent squeezed vacuum states (in the weighted superposition) are $10$ dB squeezed below vacuum variance in the $q-$quadrature, i.e., where the $|\widetilde{0}\rangle,$ $|\widetilde{1}\rangle,$ in the superposition are of the form in (\ref{finite_sq_01_states}) with $\sigma^2=0.1$ (where squeezing in dB is calculated as $-10\log_{10}\sigma^2$). 
Note that in the above descriptions of finite-energy GKP qubit states, it is implicit that the conjugate quadrature is correspondingly anti-squeezed with a variance equalling that of the Gaussian outer envelope. 
More generally, the outer Gaussian envelope of a finite-energy GKP qubit state can have a variance $\leq 1/(2\sigma^2)$, so that the squeezing-anti-squeezing product is $\leq 1/4$~\cite{MYK19}.

An equivalent, more general description of an arbitrary finite-energy GKP qubit state $|\widetilde{\psi}\rangle$ in terms of the corresponding ideal GKP qubit state $|\overline{\psi}\rangle$ (up to normalization) that was also discussed in Ref.~\cite{GKP01} is given by
\begin{align}
|\widetilde{\psi}\rangle&=\int du\, dv\, \eta(u,v)\, e^{i(-u \hat{p} + v \hat{q})}|\overline{\psi}\rangle,\label{errorwfn} \\ 
&=\int du\, dv \,\eta(u,v)e^{\frac{iuv}{2}}\, e^{-iu \hat{p}}e^{+i v\hat{q}}|\overline{\psi}\rangle,\label{errorwfn_2}\\
\eta(u,v)&=\frac{1}{\sqrt{\pi \kappa\delta}}e^{-\frac{1}{2}(\frac{(u-u')^2}{\delta^2}+\frac{(v-v')^2}{\kappa^2})};\ 0<\delta\kappa<1,\label{errorwfn_3}
\end{align}
where $\eta(u,v)$ is the square root of a real-valued, characteristic bivariate Gaussian distribution for the state, known as its error wavefunction, where $(u',v')$ and $(\delta^2,\kappa^2)$ are the means and variances of the $(q,p)-$quadrature displacements, respectively. 
The state $|\widetilde{\psi}\rangle$ is thus a coherent superposition of displaced ideal GKP qubit states, where the displacements are drawn from the distribution $|\eta(u,v)|^2$, which effectively introduces a Gaussian envelope. 
For the explicit equivalence of the error wavefuction description to the quadrature basis description, see Appendix~\ref{GKP_equivalence}. Note that a superposition of displacements is different from an incoherent mixture of displacements, which characterize thermal noise.

\section{Graph States composed of finite squeezed GKP qubits}\label{GKPgraph}
Consider a finite, simple, undirected graph $G(\mathrm{V},\mathrm{E})$, where $\mathrm{V}$ is the set of vertices $\mathrm{v}_i$ with cardinality $|\mathrm{V}|=n$, and $\mathrm{E}$ is the set of edges $\mathrm{e}_{ij}=(\mathrm{v}_i,\mathrm{v}_j)$ connecting vertices $\mathrm{v}_i,\mathrm{v}_j$. 
A graph state $|G\rangle$ is defined as
\begin{equation}
|\overline{\Psi_G}\rangle=\prod_{\mathrm{e}\in \mathrm{E}}{C_Z}_\mathrm{e}|\overline{+}\rangle^{\otimes n},
\label{ideal_GraphState}
\end{equation}
where the vertices in $\mathrm{V}$ of graph $G$ have been associated with qubits in the $|\overline{+}\rangle$ state and the edges $\mathrm{e}\in \mathrm{E}$ with the controlled-phase gate denoted by  $C_Z$.

When the vertices are bosonic modes initialized as finite-energy approximate GKP qubits of the form in (\ref{errorwfn}-\ref{errorwfn_3}), where $\delta_i^2=l_i\sigma^2,\kappa_i^2=m_i\sigma^2$ and $u_i'=\mu_{q_i},v_i'=\mu_{p_i}\forall i\in\{\mathrm{v}_1,\ldots,\mathrm{v}_n\}$, $\sigma^2$ is some unit variance and all $l_i,m_i,\mu_{q_i}\mu_{p_i}\in\mathbb{R}$, the corresponding graph state takes the form
\begin{align}
|\widetilde{\Psi_G}\rangle=\int d\vec{x}\ \eta_G(\vec{\mu},V,\vec{x})\prod_{i=1}^{n}e^{\frac{ix_ix_{n+i}}{2}}e^{-ix_i\hat{p}_i}e^{+ix_{n+i}\hat{q}_i}|\overline{\Psi_G}\rangle.\label{GKPgraphstate}
\end{align}
Here $|\overline{\Psi_G}\rangle$ denotes the ideal GKP-qubit graph state of the form in (\ref{ideal_GraphState}) associated with the graph $G(\mathrm{V,E})$, which is obtained using the continuous variable controlled-phase gate that is the quadratic Gaussian unitary given by ${C_Z}_\mathrm{e}=e^{-i\hat{q}_{\mathrm{v}_i}\hat{q}_{\mathrm{v}_j}}$ acting on edges $\mathrm{e}_{ij}=(\mathrm{v}_i,\mathrm{v}_j)$, where the vertices are associated with the ideal $|\overline{+}\rangle$ GKP qubit state of (\ref{p-GKP_Ideal}). 
In the Heisenberg picture, the $C_Z$ unitary transforms the quadrature operators of two modes $\mathrm{v}_1,\mathrm{v}_2$ symmetrically as
	\begin{align}
			C_Z^\dagger \hat{q}_{\mathrm{v}_1}C_Z&=\hat{q}_{\mathrm{v}_1}\nonumber\\
			C_Z^\dagger \hat{p}_{\mathrm{v}_1}C_Z&=\hat{p}_{\mathrm{v}_1}-\hat{q}_{\mathrm{v}_2}\nonumber\\
			C_Z^\dagger \hat{q}_{\mathrm{v}_2}C_Z&=\hat{q}_{\mathrm{v}_2}\nonumber\\
			C_Z^\dagger \hat{p}_{\mathrm{v}_2}C_Z&=\hat{p}_{\mathrm{v}_2}-\hat{q}_{\mathrm{v}_1}.
			\end{align}
The error wavefunction $\eta_{G}( \vec{\mu},V,\vec{x})$ is the square root of the $2n-$variate real-valued Gaussian distribution of the now correlated coherent random displacements $\vec{s},\vec{t}$ acting on the underlying ideal GKP qubits, given by
\begin{align}
\eta_{G}(V, \vec{\mu},\vec{x})&=\frac{e^{-\frac{1}{2}\left((\vec{x}-\vec{\mu})\cdot V^{-1}\cdot (\vec{x}-\vec{\mu})^T\right)}}{(\pi^{2n} \det(V))^{1/4}},
\end{align}
where
\begin{align}
\vec{x}&=(s_1,s_2,\ldots, s_n,t_1,t_2,\ldots, t_n),\\
\vec{\mu}&\equiv\{\langle\vec{s}\rangle,\langle\vec{t}\rangle\},\\
V&=
\begin{pmatrix}
\mathfrak{Q}_{n \times n} &\mathfrak{R}^T_{n\times n} \tabularnewline
\mathfrak{R}_{n\times n} & \mathfrak{P}_{n\times n} \tabularnewline
\end{pmatrix}\sigma^2,\label{Graph_V_structure}\\
\mathfrak{Q}_{ij}&\equiv\langle\Delta s_i \Delta s_j\rangle, \mathfrak{P}_{ij}\equiv\langle\Delta t_i \Delta t_j\rangle, \mathfrak{R}_{ij}\equiv\langle\Delta t_i \Delta s_j\rangle,
\end{align}
and the mean displacement vector $\vec{\mu}$ and the covariance matrix elements $V_{ij}$ take on the following values:
\begin{align}
\vec{\mu}&=\{\vec{\mu'}_q,\vec{\mu'}_p\},\\ 
\vec{\mu'}_q&=\{\mu_{q_i}\}_{i=1}^n,\\
\vec{\mu'}_p&=\{\mu'_{p_i}\}_{i=1}^n,\ \mu'_{p_i}=\mu_{p_i}-\sum_{j}A_{ij}\mu_{q_j},\label{graphstate_zero_mean}\\
\mathfrak{Q}&=\textrm{diag}(\vec{l}),\ \vec{l} =\{l_i\}_{i=1}^n, \\
\mathfrak{P}&=\textrm{diag}(\vec{m'}),\  \vec{m'}=\{m'_i\}_{i=1}^n,\  m'_i=m_i+\sum_{j}A_{ij}l_j , \\
\mathfrak{R}&\equiv \{\mathfrak{R}_{ij}\}_{i,j=1}^{n},\ 
\mathfrak{R}_{ij}=-A_{ij}l_j,
\end{align}
with $A$ being the adjacency matrix of graph $G(\mathrm{V,E})$. 
The above form for $\eta_G$ is the result of the action of the affine-symplectic map corresponding to the CV $C_Z$ unitary operation on the quadrature variables in phase space~\cite{WPGCRSL12}. 
The graph state of (\ref{GKPgraphstate}) most generally can thus be compactly represented by a node-weighted version of the graph $G(\mathrm{V,E})$, the node weights being the mean quadrature displacements of the modes $\vec{\mu}=(\vec{\mu'}_q,\vec{\mu'}_p)$, and in addition by specifying a $2n\times 2n$, real, symmetric covariance matrix $V$ associated with the correlated coherent random displacements $\vec{s},\vec{t}$ along the $n$ $\hat{q}$ and $\hat{p}$ quadratures, respectively, acting on the underlying ideal GKP qubit graph state. 
In other words, we can represent $|\widetilde{\Psi_G}\rangle\equiv G(\mathrm{V},\mathrm{E},\Vec{\mu}_q,\vec{\mu}_p,V)$.

Evidently, if large graph states are generated from individual finite-energy approximate GKP qubits using $C_Z$ gates alone, then the $p-$quadrature variances of the modes can quickly accumulate, rendering the state too noisy. 
The next two sections discuss tools that help remedy this situation.  

\section{GKP error correction of finite squeezed GKP qubit graph states}\label{singlequbit QEC}
The GKP Steane error correction is a procedure that helps reduce quadrature noise in the finite squeezed GKP qubits at the cost of a possible mean displacement error. 
We will begin by describing the action of the procedure on a trivial single mode graph state, i.e., a single approximate GKP qubit state of the form in (\ref{errorwfn}), and then generalize the results to a vertex of a larger graph. 
The procedure involves a unitary interaction between the ``data'' qubit, whose noise along $p$ or $q$ quadrature is desired to be reduced, and an ancilla that is also prepared in a finite-energy approximation of an eigenstate of the conjugate quadrature, but with presumably lower noise variance than the data qubit in the quadrature of interest. 
We note that Steane error correction for the case of a single qubit graph has been previously discussed in \cite{GK06}.

\begin{figure}
    \centering
    \includegraphics[width=0.25\textwidth]{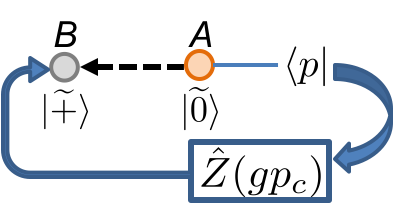}
    \caption{A schematic of the Steane error correction procedure for reducing the noise variance along the $p-$quadruture in an approximate GKP qubit using an approximate GKP ancilla qubit. The black dashed arrow indicates a Controlled-Not gate.}
    \label{fig:pSteane}
\end{figure}

We discuss GKP-Steane error correction for $p-$quadrature noise reduction here. 
A schematic of the procedure is shown in Figure~\ref{fig:pSteane}. 
It consists of preparing an ancilla GKP qubit (qubit $A$) in the $|\widetilde{0}\rangle$ state of (\ref{q-GKP_Ideal}) and performing the CV Controlled-NOT gate $C_X$ with the ancilla qubit as the control and the data qubit (qubit $B$) as the target. 
This is followed by performing a $p-$quadrature homodyne measurement on the ancilla and a feedback $\hat{Z}(g y)$ displacement on the data GKP qubit, where $y$ is the measurement outcome and $g$ is a suitable gain factor. 
In this work, the feedback displacement is chosen such that it removes the measurement outcome-dependent component of the conditional mean displacement on the data qubit(s). 
The $C_X$ gate in CV is given by the unitary interaction $C_X^{C\rightarrow T}=e^{-i\hat{q}_C\hat{p}_T},$ where $C,T$ denote the control and target modes, respectively. 
In the Heisenberg picture, the $C_X$ unitary transforms the quadrature operators of control and target modes as
	\begin{align}
			C_X^\dagger \hat{q}_{C}C_X&=\hat{q}_{C}\nonumber\\
			C_X^\dagger \hat{p}_{C}C_X&=\hat{p}_{C}-\hat{p}_{T}\nonumber\\
			C_X^\dagger \hat{q}_{T}C_X&=\hat{q}_{T}+\hat{q}_{C}\nonumber\\
			C_X^\dagger \hat{p}_{T}C_X&=\hat{p}_{T}.
	\end{align}
The CV $C_X$ gate can be implemented using beam splitters and inline single-mode squeezers as described later in Sec.~\ref{fusion_ops}.

Note that the procedure for $q-$quadrature noise reduction would similarly involve an ancilla prepared in a $|\widetilde{+}\rangle$ and the $C_X$ between the data and the ancilla qubits, but with the data qubit as the control and the ancilla qubit as the target. 
This would be followed by a $q-$quadrature measurement of the ancilla qubit and a feedback measurement on the data qubit.

\subsection{$p-$Steane error correction of a single finite-energy GKP qubit}
In $p-$quadrature GKP Steane error correction of a single finite-energy GKP qubit in a state of the form in (\ref{errorwfn}), the interaction between the data and ancilla approximate GKP qubits results in a two-mode state given by 
\begin{align}
    &|\psi\rangle_{AB}=C_X^{A\rightarrow B}|\widetilde{0}\rangle_A\otimes|\widetilde{+}\rangle_B\nonumber\\
	&=\int du_A \,dv_A \,\eta_A(u_A,v_A)\int du_B\,dv_B\,\eta_B(u_B,v_B)\nonumber\\
	&\times C_X^{A\rightarrow B}e^{i(-u_A\hat{p}_A+v_A\hat{q}_A)}e^{i(-u_B\hat{p}_B+v_B\hat{q}_B)}|\overline{0}\rangle_A|\overline{+}\rangle_B\\
	&=\int ds_A\,dt_A\, ds_B\,dt_B\,\eta_{AB}(s_A,s_B,t_A,t_B)\nonumber\\
	&\times e^{i(-s_A\hat{p}_A+t_A\hat{q}_A)}e^{i(-s_B\hat{p}_B+t_B\hat{q}_B)}|\overline{0}\rangle_A|\overline{+}\rangle_B,
\end{align}
where $s_A=u_A,\ s_B=u_B+u_A,\ t_A= v_A-v_B,\ t_B=v_B$ and $\eta_{AB}(s_A,s_B,t_A,t_B)$ is the square root of a 4-variate Gaussian distribution
\begin{align}
	\eta_{AB}(s_A,s_B,t_A,t_B)&=\frac{e^{-\frac{1}{2}\left(\vec{x}V^{-1}\vec{x}^T\right)}}{\pi (\det(V))^{1/4}},
\end{align}
with
\begin{align}
	\vec{x}&=(s_A,s_B,t_A,t_B),\\
	V&=\mathfrak{Q}\oplus \mathfrak{P},\\
	\mathfrak{Q}=\begin{pmatrix}
	l_A& l_A\tabularnewline
	l_A & l_B+l_A \tabularnewline
	\end{pmatrix}&\sigma^2,\ 
	\mathfrak{P}=\begin{pmatrix}
	m_A+m_B &-m_B \tabularnewline
	-m_B &m_B \tabularnewline
	\end{pmatrix}\sigma^2\label{Steane_vertex_lm_values}
\end{align}
We note that $C_X^{A\rightarrow B}|\overline{0}\rangle_A\otimes|\overline{+}\rangle_B=|\overline{0}\rangle_A\otimes|\overline{+}\rangle_B.$
	
The state $|\psi\rangle_{AB}$ can be equivalently written as~\cite{Wang2019}
\begin{align}
&\int ds_Ads_B \chi_{AB}(s_A,s_B)\int dt_Adt_B\mathcal{P}(t_A)\mathcal{Q}(t_B|t_A)\nonumber\\
&\times e^{i(-s_A\hat{p}_A+t_A\hat{q}_A)}e^{i(-s_B\hat{p}_B+t_B\hat{q}_B)}|\overline{0}\rangle_A|\overline{+}\rangle_B,
\end{align}
where $\chi_{AB}(s_A,s_B)$, and $\{\mathcal{P}(t_A),\ \mathcal{Q}(t_B|t_A)\}$ are the square roots of bivariate and univariate, real-valued Gaussian distributions, respectively,  denoted as
\begin{align}
\chi_{AB}(s_A,s_B)&\sim\mathcal{G}_\mathbb{R}^{1/2}((0,0),\mathfrak{Q}),\\
\mathcal{P}(t_A)&\sim\mathcal{G}_\mathbb{R}^{1/2}(0,(m_A+m_B)\sigma^2),\\
\mathcal{Q}(t_B|t_A)&\sim\mathcal{G}_\mathbb{R}^{1/2}\left(\frac{-m_B }{m_A+m_B}t_A,\frac{m_Am_B}{m_A+m_B}\sigma^2\right).\label{Steane_distributions}
\end{align}
The latter follows by applying Bayes' rule.
	
\subsubsection*{Ancilla measurement and conditional post-measurement state} %\noindent\textbf{Ancilla measurement and conditional post-measurement state:} 
When qubit $A$ is measured over its $p-$quadrature, we get an outcome $y\in\mathbb{R}$ with probability $P_{Y}(y)$ and a conditional state $|\phi\rangle_{B|y}$ on qubit B, that can be deduced from the conditional unnormalized state given by (see Appendix~\ref{Steane_EC_details} for the derivation)
\begin{align}
&_p\langle y|_A.|\psi\rangle_{AB}=\sqrt{P_{Y}(y)}|\phi\rangle_{B|y}\nonumber\\
&=2\sqrt{\pi}\sum_n\frac{e^{-\frac{(y-n\sqrt{\pi})^2}{2(m_A+m_B)\sigma^2}}}{(\pi(m_A+m_B)\sigma^2)^{1/4}}\frac{e^{-\frac{(y+n\sqrt{\pi})^2}{8\frac{(l_A+l_B)}{l_Al_B\sigma^2}}}}{(4\pi\frac{(l_A+l_B)}{l_Al_B\sigma^2})^{1/4}}\nonumber\\
&\times\Bigg(\int ds_B\,dt_B\, \frac{e^{-\frac{s_B^2}{2(l_A+l_B)\sigma^2}}}{(\pi(l_A+l_B)\sigma^2)^{1/4}}\frac{e^{-\frac{\left(t_B+\frac{m_B}{m_A+m_B}(y-n\sqrt{\pi})\right)^2}{2\frac{m_Am_B}{m_A+m_B}\sigma^2}}}{(\pi\frac{m_Am_B}{m_A+m_B}\sigma^2)^{1/4}}\nonumber\\
	&e^{-i\frac{l_A}{2(l_A+l_B)}(y+n\sqrt{\pi})s_B}e^{i(-s_B\hat{p}_B+t_B\hat{q}_B)}|\overline{+}\rangle_B\Bigg)
\end{align}
Assuming $\sqrt{(m_A+m_B)}\sigma\ll \sqrt{\pi}/2$, the support of $P_{Y}(y)$ is mainly concentrated around $y=n\sqrt{\pi},\ n\in\mathbb{Z}$. When $y-n\sqrt{\pi}$ is small for some $n\in\mathbb{Z}$, then $y+n\sqrt{\pi}\approx 2n\sqrt{\pi}$ for that $n$, i.e., $\sqrt{P_{Y}(y)}|\phi\rangle_{B|y}$
\begin{align}
&\propto\sum_n\frac{e^{-\frac{(2n\sqrt{\pi})^2}{8\frac{(l_A+l_B)}{l_Al_B\sigma^2}}}}{(4\pi\frac{(l_A+l_B)}{l_Al_B\sigma^2})^{1/4}}\frac{e^{-\frac{(y-n\sqrt{\pi})^2}{2(m_A+m_B)\sigma^2}}}{(\pi(m_A+m_B)\sigma^2)^{1/4}}\nonumber\\
&\Bigg(\int ds_B\,dt_B\, \frac{e^{-\frac{s_B^2}{2(l_A+l_B)\sigma^2}}}{(\pi(l_A+l_B)\sigma^2)^{1/4}}\frac{e^{-\frac{\left(t_B+\frac{m_B}{m_A+m_B}(y-n\sqrt{\pi})\right)^2}{2\frac{m_Am_B}{m_A+m_B}\sigma^2}}}{(\pi\frac{m_Am_B}{m_A+m_B}\sigma^2)^{1/4}}\nonumber\\
&e^{-i\frac{l_A}{l_A+l_B}n\sqrt{\pi}s_B}e^{i(-s_B\hat{p}_B+t_B\hat{q}_B)}|\overline{+}\rangle_B\Bigg).
\end{align}
Moreover, $P_{Y}(y)$  can be approximated as
\begin{align}
&\propto\sum_n \left|\frac{e^{-\frac{(n\sqrt{\pi})^2}{2\frac{(l_A+l_B)}{l_Al_B\sigma^2}}}}{(\pi\frac{(l_A+l_B)}{l_Al_B\sigma^2})^{1/4}}\frac{e^{-\frac{(y-n\sqrt{\pi})^2}{2(m_A+m_B)\sigma^2}}}{(\pi(m_A+m_B)\sigma^2)^{1/4}}\right|^2\\
&=\sum_n \frac{e^{-\frac{(n\sqrt{\pi})^2}{\frac{(l_A+l_B)}{l_Al_B\sigma^2}}}}{(\pi\frac{(l_A+l_B)}{l_Al_B\sigma^2})^{1/2}}\frac{e^{-\frac{(y-n\sqrt{\pi})^2}{(m_A+m_B)\sigma^2}}}{(\pi(m_A+m_B)\sigma^2)^{1/2}}\\
&=\sum_nP_N[n]P_Q(y-n\sqrt{\pi}),\label{Steane Py}
\end{align}
where the random variables $N$ and $Q$ are given by
\begin{align}
    N&\sim\mathcal{G}_\mathbb{Z}(0,\left(\frac{l_A+l_B}{l_Al_B}\right)/(2\pi\sigma^2)),\ n\in\mathbb{Z},\label{Steane_PN}\\
    Q&\sim\mathcal{G}_\mathbb{R}(0,(m_A+m_B)\sigma^2/2),\ p\in\mathbb{R}.\label{Steane_PQ}
\end{align}
That is, the outcome of the $p-$homodyne measurement of qubit $A$ is
\begin{align}
    y=\sqrt{\pi} N+Q,
\end{align}
and its distribution is given by the convolution of $N\sqrt{\pi}\sim\mathcal{G}_\mathbb{R}(0,\left(\frac{l_A+l_B}{l_Al_B}\right)/(2\sigma^2)),\ n\in\mathbb{Z}$ and $Q$.
Note that $Q$ has the same distribution as $t_A$. 

Thus, the conditional post-measurement state of qubit $B$ can be written as
\begin{align}
|\phi\rangle_{B|y}&=\frac{1}{\sqrt{\mathcal{N}}}\sum_n\sqrt{P_N[n]P_Q(y-n\sqrt{\pi})}\int ds_Bdt_B \nonumber\\
&\zeta(s_B,t_B)e^{-i\frac{l_A}{l_A+l_B}n\sqrt{\pi}s_B}e^{i(-s_B\hat{p}_B+t_B\hat{q}_B)}|\overline{+}\rangle_B,\\
\zeta(s_B,t_B)&=\frac{e^{-\frac{s_B^2}{2(l_A+l_B)\sigma^2}}}{(\pi(l_A+l_B)\sigma^2)^{1/4}}\frac{e^{-\frac{\left(t_B+\frac{m_B}{m_A+m_B}(y-n\sqrt{\pi})\right)^2}{2\frac{m_Am_B}{m_A+m_B}\sigma^2}}}{(\pi\frac{m_Am_B}{m_A+m_B}\sigma^2)^{1/4}},\\
\mathcal{N}&=\sum_nP_N[n]P_Q(y-n\sqrt{\pi})\label{vertex p Steane norm}.
\end{align}
By a change of variables $t_B\rightarrow t_B-\frac{m_B}{m_A+m_B}(y-n\sqrt{\pi})$, the above state can be equivalently written as
\begin{align}
|\phi\rangle_{B|y}&=\frac{1}{\sqrt{\mathcal{N}}}\sum_n\sqrt{P_N[n]P_Q(y-n\sqrt{\pi})}\nonumber\\
&\int ds_B\,dt_B\, \zeta(s_B,t_B)e^{-i\frac{l_A}{l_A+l_B}n\sqrt{\pi}s_B}\nonumber\\
&e^{i(-s_B\hat{p}_B+\left(t_B-\frac{m_B}{m_A+m_B}(y-n\sqrt{\pi})\right)\hat{q}_B)}|\overline{+}\rangle_B,\\
\zeta(s_B,t_B)&=\frac{e^{-\frac{s_B^2}{2(l_A+l_B)\sigma^2}}}{(\pi(l_A+l_B)\sigma^2)^{1/4}}\frac{e^{-\frac{t_B^2}{2\frac{m_Am_B}{m_A+m_B}\sigma^2}}}{(\pi\frac{m_Am_B}{m_A+m_B}\sigma^2)^{1/4}},
\end{align}
where $\mathcal{N}$ is the normalization factor of (\ref{vertex p Steane norm}) and the mean displacement has been moved to the displacement operator from the error wavefunction.

\subsubsection*{Feedback displacement on $B$}
%\noindent\textbf{Feedback displacement on $B$:} 
Following the ancilla measurement, a feedback displacement $\hat{Z}(gp_c(y))$, $y$ being the ancilla measurement outcome and $g$ being a gain factor, is applied on mode $B$. The quantity $p_c(y)$ is chosen to the amount of
\begin{align}
    p_c(y)&=y\mod\sqrt{\pi}
\end{align}
such that $p_c(y)\in[-\sqrt{\pi}/2,\sqrt{\pi}/2]$.
%\Prajit{I don't think the meaning of (53) is clear.}
That is, e.g.,
\begin{align}
    0<y<\frac{\sqrt{\pi}}{2}&\Rightarrow p_c(y)=y\\
    \frac{\sqrt{\pi}}{2}<y<\sqrt{\pi}&\Rightarrow p_c(y)=y-\sqrt{\pi}.
\end{align}
More generally, for $y>0$ and $n'=\floor{\frac{y}{\sqrt{\pi}}}\in\mathbb{Z}$, we have $p_c(y)$
\begin{align}
&= \begin{cases}
y-n'\sqrt{\pi},            & y-n'\sqrt{\pi} <\sqrt{\pi}/2\\
y-(n'+1)\sqrt{\pi},               &\sqrt{\pi}> y-n'\sqrt{\pi} >\sqrt{\pi}/2
\end{cases}.
\end{align}
Likewise, for $y<0$ and $n'=\floor{\frac{|y|}{\sqrt{\pi}}}\in\mathbb{Z}$, we have $p_c(y)$
\begin{align}
&= \begin{cases}
y+n'\sqrt{\pi},            & |y|-n'\sqrt{\pi} <\sqrt{\pi}/2\\
y+(n'+1)\sqrt{\pi},               &\sqrt{\pi}>|y|-n'\sqrt{\pi} >\sqrt{\pi}/2
\end{cases}.
\end{align}
When the gain factor is chosen to be $g=\frac{m_B}{m_A+m_B}$, for an ancilla measurement outcome in the interval of $0<y<\frac{\sqrt{\pi}}{2}$, the state of qubit $B$ is transformed as
$|\phi\rangle_{B|y}\rightarrow |\psi\rangle_{B|y}$, where $|\psi\rangle_{B|y}$
\begin{align}
&=\hat{Z}\left(\frac{m_B}{m_A+m_B}p_c(y)\right)|\phi\rangle_{B|y}\nonumber\\
&=\frac{1}{\sqrt{\mathcal{N}}}\sum_n\sqrt{P_N[n]P_Q(y-n\sqrt{\pi})}\nonumber\\
&\int ds_B\,dt_B\, \zeta(s_B,t_B)\, e^{is_B\left(\frac{m_B}{m_A+m_B}\frac{y}{2}-\frac{l_A}{l_A+l_B}n\sqrt{\pi}\right)}\nonumber\\
&e^{(i(-s_B\hat{p}_B+\left(t_B+\frac{m_B}{m_A+m_B}n\sqrt{\pi}\right)\hat{q}_B)}|\overline{+}\rangle_B,
\end{align}
$\mathcal{N}$ being the normalization factor of (\ref{vertex p Steane norm}). 
Note that with this choice of $g$, we get rid of the $y-$dependent mean displacement. 
On the contrary, there is now a $y-$dependent phase factor, which is arrived at by carefully manipulating the joint displacement operator with the $\hat{Z}$ displacement from the feedback using the decompositions in (\ref{XZ_ZX_displ}). 
However, this phase is inconsequential since it is a global phase, present in all terms in the coherent superposition. 
When $m_A\ll m_B$, we have $|\psi\rangle_{B|y}$
\begin{align}
&\approx\frac{1}{\sqrt{\mathcal{N}}}\sum_n\sqrt{P_N[n]P_Q(y-n\sqrt{\pi})}\int ds_B\,dt_B\, \xi(s_B,t_B)\nonumber\\
& e^{is_B\left(\frac{y}{2}-\frac{l_A}{l_A+l_B}n\sqrt{\pi}\right)}e^{i(-s_B\hat{p}_B+\left(t_B+n\sqrt{\pi}\right)\hat{q}_B)}|\overline{+}\rangle_B,
\end{align}
where
\begin{align}
\xi(s_B,t_B)&=\lim_{m_A/m_B\rightarrow 0}\zeta(s_B,t_B)\nonumber\\
&=\frac{e^{-\frac{s_B^2}{2(l_A+l_B)\sigma^2}}}{(\pi(l_A+l_B)\sigma^2)^{1/4}}\frac{e^{-\frac{t_B^2}{2m_A\sigma^2}}}{(\pi m_A\sigma^2)^{1/4}}\label{errorwfnSteane}.
\end{align}

More generally, when $m_A\ll m_B$, for any $y \gtrless 0$, we have $|\psi\rangle_{B|y}\approx$
\begin{align}
&\frac{1}{\sqrt{\mathcal{N}}}\sum_n\sqrt{P_N[n]P_Q(y-n\sqrt{\pi})}\int ds_B\,dt_B\, \xi(s_B,t_B)\nonumber\\
& e^{is_B\left(\frac{y\mp z\sqrt{\pi}}{2}-\frac{l_A}{l_A+l_B}n\sqrt{\pi}\right)}e^{i(-s_B\hat{p}_B+\left(t_B+(n\mp z)\sqrt{\pi})\right)\hat{q}_B}|\overline{+}\rangle_B,\\
&z = \begin{cases}
n',            & |y|-n'\sqrt{\pi} <\sqrt{\pi}/2\\
(n'+1),               &\sqrt{\pi}>|y|-n'\sqrt{\pi} >\sqrt{\pi}/2
\end{cases},\\
&n'=\floor{\frac{|y|}{\sqrt{\pi}}}\in\mathbb{Z},
\end{align}
where $\xi(s_B,t_B)$ is as given in (\ref{errorwfnSteane}).

\subsubsection*{Mean displacement error in the Steane error-corrected state}
%\noindent\textbf{Mean displacement error in the Steane error-corrected state:}\\
We will now focus on the interval $|y|<\sqrt{\pi}$ for the ancilla measurement outcome. 
Once again, assuming $\sqrt{(m_A+m_B)}\sigma\ll \sqrt{\pi}/2$, the probability $P_Y(y)$ of (\ref{Steane Py}) for outcome $y$ can be approximated as
\begin{align}
P_{Y}(y)\approx P_N[0]P_Q(y)+P_N[1]P_Q(\sqrt{\pi}-|y|).
\label{Steane PYy}
\end{align}
Further, when $|y|\leq \sqrt{\pi}/2$ and $m_A\ll m_B$, the error-corrected state can be approximated as $|\psi\rangle_{B|y}\approx$
\begin{align}
&\frac{\sqrt{P_N[0]P_Q(|y|)}|\psi_0\rangle+\sqrt{P_N[1]P_Q(\sqrt{\pi}-|y|)}|\psi_1\rangle}{\sqrt{P_Y(y)}},
\end{align}
where
\begin{align}
|\psi_0\rangle&=\int ds_B\,dt_B\, \xi(s_B,t_B)\,e^{i(-s_B\hat{p}_B+t_B\hat{q}_B)}|\overline{+}\rangle_B,\label{gplusSteane0}\\
|\psi_1\rangle&=\int ds_B\,dt_B\, \xi(s_B,t_B)\,e^{i(-s_B\hat{p}_B+\left(t_B-\sqrt{\pi}\right)\hat{q}_B)}|\overline{+}\rangle_B,\nonumber\\
&=\int ds_B\,dt_B\, \xi(s_B,t_B+\sqrt{\pi})\,e^{i(-s_B\hat{p}_B+t_B\hat{q}_B)}|\overline{+}\rangle_B,\label{gplusSteane1}
\end{align}
and $P_Y(y)$ and $\xi(s_B,t_B)$ are as given in (\ref{Steane PYy}) and (\ref{errorwfnSteane}), respectively.
Note that we are ignoring phase factors in the above expressions since they do not affect the mean displacement error probabilities discussed below.

We observe that the $p-$displacements in $|\psi_1\rangle$ have an additional mean shift of $\sqrt{\pi}$, i.e., an offset of half the GKP grid spacing relative to $|\psi_0\rangle$. 
Since the underlying ideal GKP qubit state in both $|\psi_0\rangle$ and $|\psi_1\rangle$ is an eigenstate of the $\hat{X}$ operator (the $|\overline{+}\rangle$ state), the additional mean $p-$displacement of $\sqrt{\pi}$ in $|\psi_1\rangle$ implies a logical $Z-$flip of the underlying ideal GKP qubit state. 
In other words, the $|\psi_1\rangle$ has an orthogonal support compared to $|\psi_0\rangle$ in the GKP grid state basis.

However, it should be noted that $|\psi_1\rangle$ is not exactly the zero mean finite squeezed $|\widetilde{-}\rangle$, because it has its Gaussian envelope offset relative to that state. 
The $p-$quadrature wavefunction of the state (and more generally for states corresponding to shifted-mean error wavefunctions) is derived in Appendix~\ref{logical_error_in_shifted_efn}. 
Figure~\ref{fig:logical_error_ewfn} plots the $p-$quadrature wavefunction of $|\psi_1\rangle$ at the output of the $p-$quadrature Steane error correction when starting with an input-ancilla error wavefunction covariance matrix of the form in (\ref{Steane_vertex_lm_values}) with $l_A=m_A=l_B=1,m_B\gg 1$ and $\sigma^2=0.1$ (corresponding to 10 dB vacuum squeezing along the $p-$quadrature). 
This output $|\psi_1\rangle$ has its error wavefunction covariance matrix specified by $\delta^2=2\sigma^2, \kappa^2=\sigma^2$. 
Figure~\ref{fig:logical_error_ewfn} also contrasts the plot of $|\psi_1\rangle$ with that of the $|\widetilde{-}\rangle$ corresponding to the same error wavefunction covariance matrix state. %The wavefunctions are plotted for the output of $p-$quadrature Steane error correction on a state $|\widetilde{+}\rangle$ of the form in (\ref{}), with $\delta^2=\sigma^2, \kappa^2=2\sigma^2$, using an ancilla $|\widetilde{0}\rangle$ state with $\delta^2=\sigma^2, \kappa^2=\sigma^2$. The output is a state of the form in (\ref{errorwfn}-\ref{errorwfn_3}) with $\delta^2/2=\kappa^2=\sigma^2=0.25$, where the latter corresponds to $6$ dB squeezing in the $p-$quadrature.
Since $|y|\leq \sqrt{\pi}/2$, we have $P_N[0]P_Q(|y|)>P_N[1]P_Q(\sqrt{\pi}-|y|)$, which implies a higher weight for $|\psi_0\rangle$ in the superposition.
\begin{figure}
    \centering
    \includegraphics[width=0.45\textwidth]{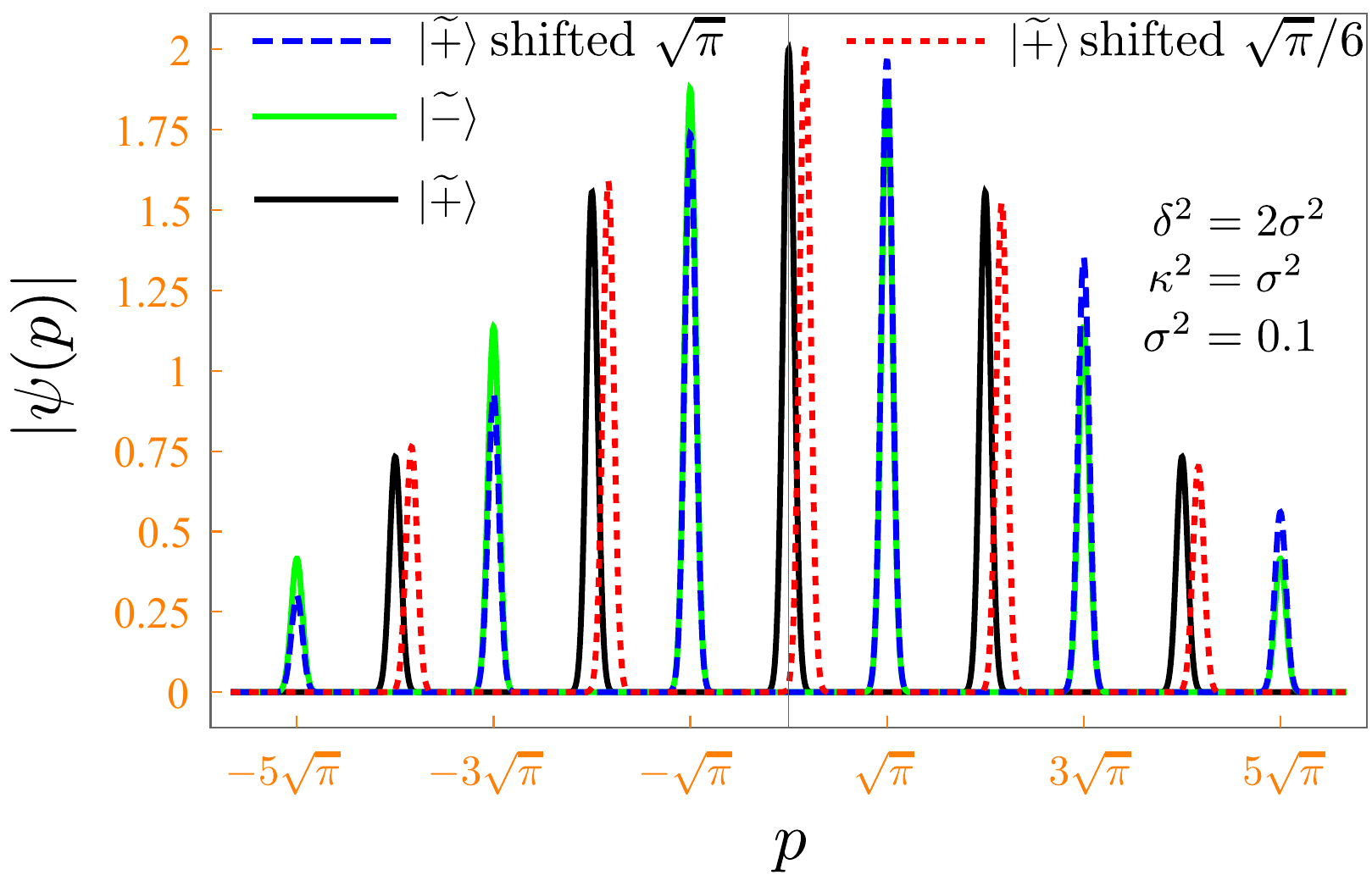}
    \caption{$p-$quadrature wavefunctions of $|\widetilde{\pm}\rangle$ states of the form in (\ref{errorwfn}-\ref{errorwfn_3}) with $\delta^2/2=\kappa^2=\sigma^2=0.1$ (corresponds to $10$ dB $p-$quadrature squeezing), and with and without mean shifts in the error wavefunction.}
    \label{fig:logical_error_ewfn}
\end{figure}
\begin{figure}
    \centering
    \includegraphics[width=0.44\textwidth]{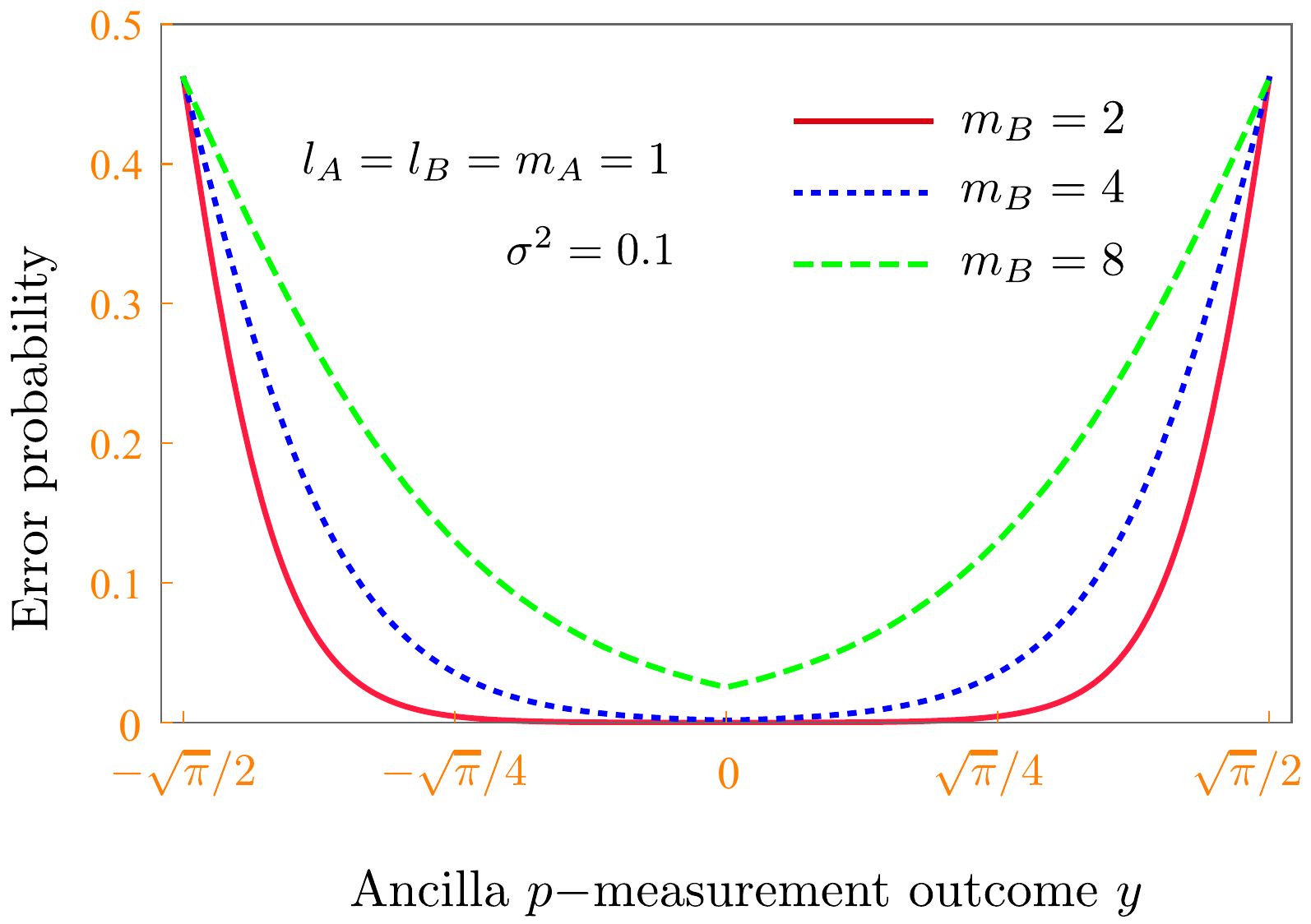}
    \caption{Mean displacement error probability as a function of the homodyne measurement outcome $|y|\leq \sqrt{\pi}/2$ for different values of $m_B$ in (\ref{Steane_vertex_lm_values}), with $l_A=l_B=m_A=1$ and $\sigma^2=0.1$.}
    \label{fig:errorprob}
\end{figure}
\begin{figure}
    \centering
    \includegraphics[width=0.44\textwidth]{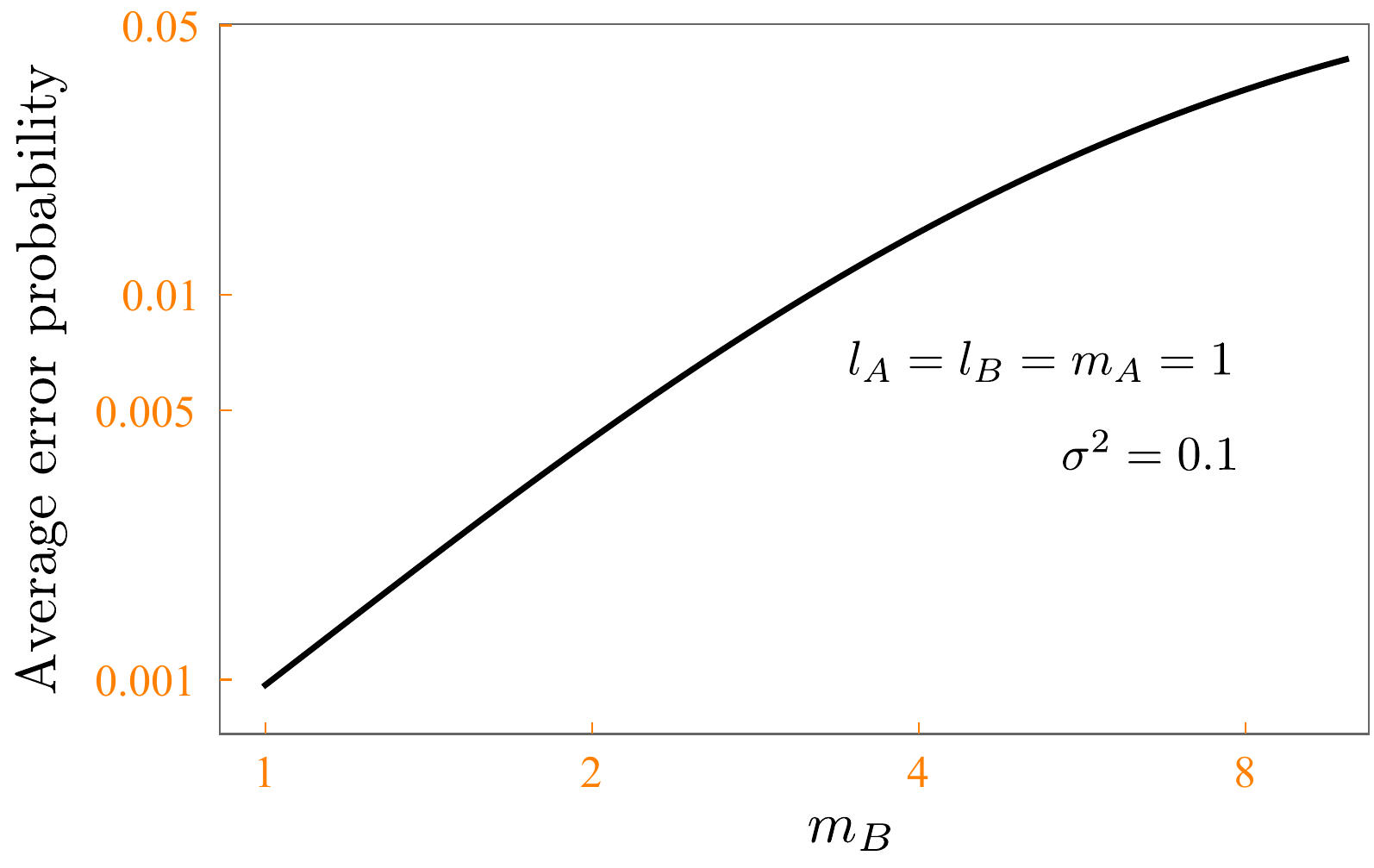}
    \caption{Average error probability when $|y|\leq \sqrt{\pi}/2$, as a function of $m_B$ in (\ref{Steane_vertex_lm_values}), with $l_A=l_B=m_A=1$ and $\sigma^2=0.1$.}
    \label{fig:errorprob2}
\end{figure}

On the other hand, when $\sqrt{\pi}>|y|> \sqrt{\pi}/2$ and $m_A\ll m_B$, we have the conditional state being $|\psi\rangle_{B|y}\approx$
\begin{align}
\frac{\sqrt{P_N[0]P_Q(|y|)}|\psi_0\rangle+\sqrt{P_N[1]P_Q(\sqrt{\pi}-|y|)}|\psi_1\rangle}{\sqrt{P_N[0]P_Q(|y|)+P_N[1]P_Q(\sqrt{\pi}-|y|)}},
\end{align}
where
\begin{align}
|\psi_0\rangle&=\int ds_B\,dt_B\, \xi(s_B,t_B)\,e^{i(-s_B\hat{p}_B+\left(t_B\mp\sqrt{\pi}\right)\hat{q}_B)}|\overline{+}\rangle_B \nonumber\\
&=\int ds_B\,dt_B\, \xi(s_B,t_B\pm\sqrt{\pi})\,e^{i(-s_B\hat{p}_B+t_B\hat{q}_B)}|\overline{+}\rangle_B,
\end{align}
and $|\psi_1\rangle$
\begin{align}
&=\int ds_B\,dt_B\, \xi(s_B,t_B)\,e^{i(-s_B\hat{p}_B+\left(t_B+(1\mp1)\sqrt{\pi}\right)\hat{q}_B)}|\overline{+}\rangle_B\nonumber\\
&=\int ds_B\,dt_B\, \xi(s_B,t_B-(1\mp1)\sqrt{\pi})\,e^{i(-s_B\hat{p}_B+t_B\hat{q}_B)}|\overline{+}\rangle_B,
\end{align}
where the $|\psi_1(y)\rangle$ now is the state with support on the original underlying ideal GKP qubit state, whereas $|\psi_0(y)\rangle$ is the state with the orthogonal support. 
The $|\psi_1(y)\rangle$ term dominates in the superposition when $P_N[1]P_Q(\sqrt{\pi}-|y|)$ is larger than $P_N[0]P_Q(|y|)$.

More generally, consider the interval of measurement outcomes given by $n\sqrt{\pi}<|y|<(n+1)\sqrt{\pi}$. The probability $P_Y(y)$ of (\ref{Steane Py}) for such an outcome $y\gtrless 0$ can be best approximated as
\begin{align}
P_{Y}(y)&\approx P_N[n]P_Q(|y|-n\sqrt{\pi})\nonumber\\
&+P_N[(n+1)]P_Q((n+1)\sqrt{\pi}-|y|).
\end{align}
When $m_A\ll m_B$, the error-corrected state is given by
\begin{align}
|\psi\rangle_{B|y}&\approx \frac{\sqrt{c_n}|\psi_n\rangle+\sqrt{c_{n+1}}|\psi_{n+1}\rangle}{\sqrt{c_n+c_{n+1}}}, \\
|\psi_n\rangle&=\int ds_B\,dt_B\, \xi(s_B,t_B)\nonumber\\
&e^{i(-s_B\hat{p}_B+\left(t_B+(n\mp z)\sqrt{\pi}\right)\hat{q}_B)}|+\rangle_B\\
|\psi_{n+1}\rangle&=\int ds_B\,dt_B\, \xi(s_B,t_B)\nonumber\\
&e^{i(-s_B\hat{p}_B+\left(t_B+(n+1\mp z)\sqrt{\pi}\right)\hat{q}_B)}|+\rangle_B\\
c_n&=P_N[n]P_Q(|y|-n\sqrt{\pi})\label{eq:cn}\\
c_{n+1}&=P_N[(n+1)]P_Q((n+1)\sqrt{\pi}-|y|),\label{eq:cn+1}\\
z&=\begin{cases}
n\sqrt{\pi},            & |y|-n\sqrt{\pi} <\sqrt{\pi}/2\\
(n+1)\sqrt{\pi},               &\sqrt{\pi}>|y|-n\sqrt{\pi} >\sqrt{\pi}/2
\end{cases}.
\end{align}
The mean displacement error probability as a function of the measurement outcome $y$ is thus given by
\begin{equation}
P_b(y) = \begin{cases}
c_{n+1}/(c_n+c_{n+1}),               & |y|-n\sqrt{\pi} <\sqrt{\pi}/2\\
c_n/(c_n+c_{n+1}),                & \sqrt{\pi}>|y|-n\sqrt{\pi} >\sqrt{\pi}/2
\end{cases}.
\label{Steane_logical_error_prob}
\end{equation}
Figure~\ref{fig:errorprob} plots the error probablity as a function of the homodyne outcome for $|y|\leq \sqrt{\pi}/2$ and different values of $m_B$ in (\ref{Steane_vertex_lm_values}), with $l_A=l_B=m_A=1$ and $\sigma^2=0.1$, while Fig.~\ref{fig:errorprob2} plots the average value of the mean-displacement error probability for $|y|\leq \sqrt{\pi}/2$ (calculated as $\int dy P_Y(y)P_b(y)$) as a function of $m_B$. 

\subsubsection*{Post-selection to minimize mean displacement error probability} 
%\noindent\textbf{Post-selection to minimize logical flip probability:} 
Consider the case $|y|<\sqrt{\pi}$. 
The post-measurement state can be enhanced by suppressing the logically-flipped component in the coherent superposition in the conditional post-measurement state. 
This can be achieved by discarding outcomes in the interval $\sqrt{\pi}/2-\nu\leq |y|\leq \sqrt{\pi}/2+\nu,\ \sqrt{\pi}/2>\nu>0$~\cite{Fukui2020-hx, FTOF18} for the following reason. 
When $|y|\leq \sqrt{\pi}/2-\nu$, the post-measurement state has $P_N[0]P_Q(|y|)/\left(P_N[1]P_Q(\sqrt{\pi}-|y|)\right)>1$ that increases with increasing $\nu$---i.e., the state $|\psi\rangle_{B|y}\approx |\psi_0(y)\rangle$ as $\nu$ increases. 
Likewise, when $\sqrt{\pi}>|y|\geq \sqrt{\pi}/2+\nu$, the post-measurement state $|\psi\rangle_{B|y}\approx |\psi_1(y)\rangle$, increasingly so, as $\nu$ increases.

\begin{figure}
    \centering
    \includegraphics[width=0.44\textwidth]{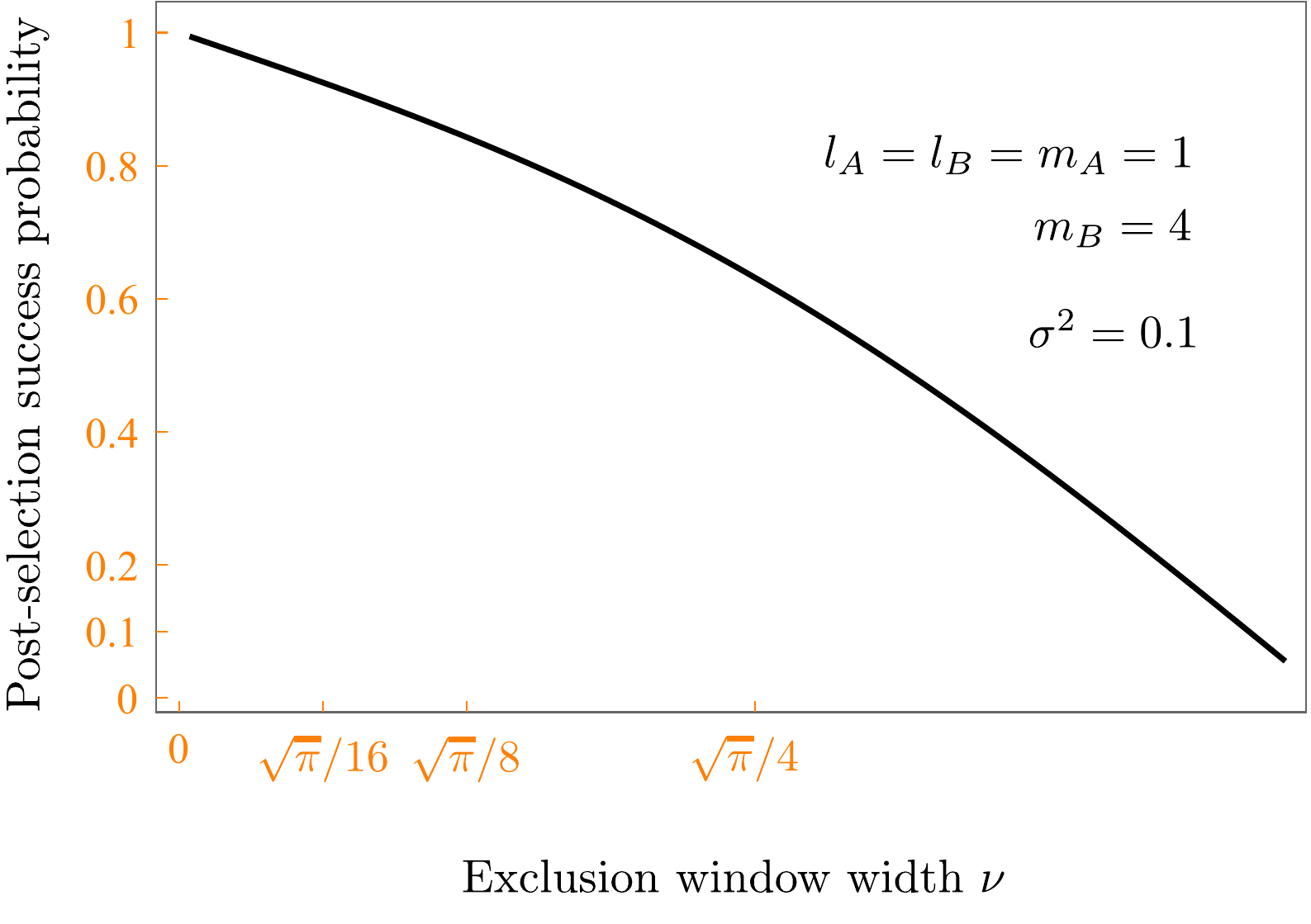}
    \caption{Post-selection success probability as a function of the exclusion window width $\nu$ in (\ref{psucc_vs_nu}), for  $l_A=l_B=m_A=1$, $m_B=4$ and $\sigma^2=0.1$ in (\ref{Steane_vertex_lm_values}).}
    \label{fig:psucc_vs_nu}
\end{figure}

\begin{figure}
    \centering
    \includegraphics[width=0.44\textwidth]{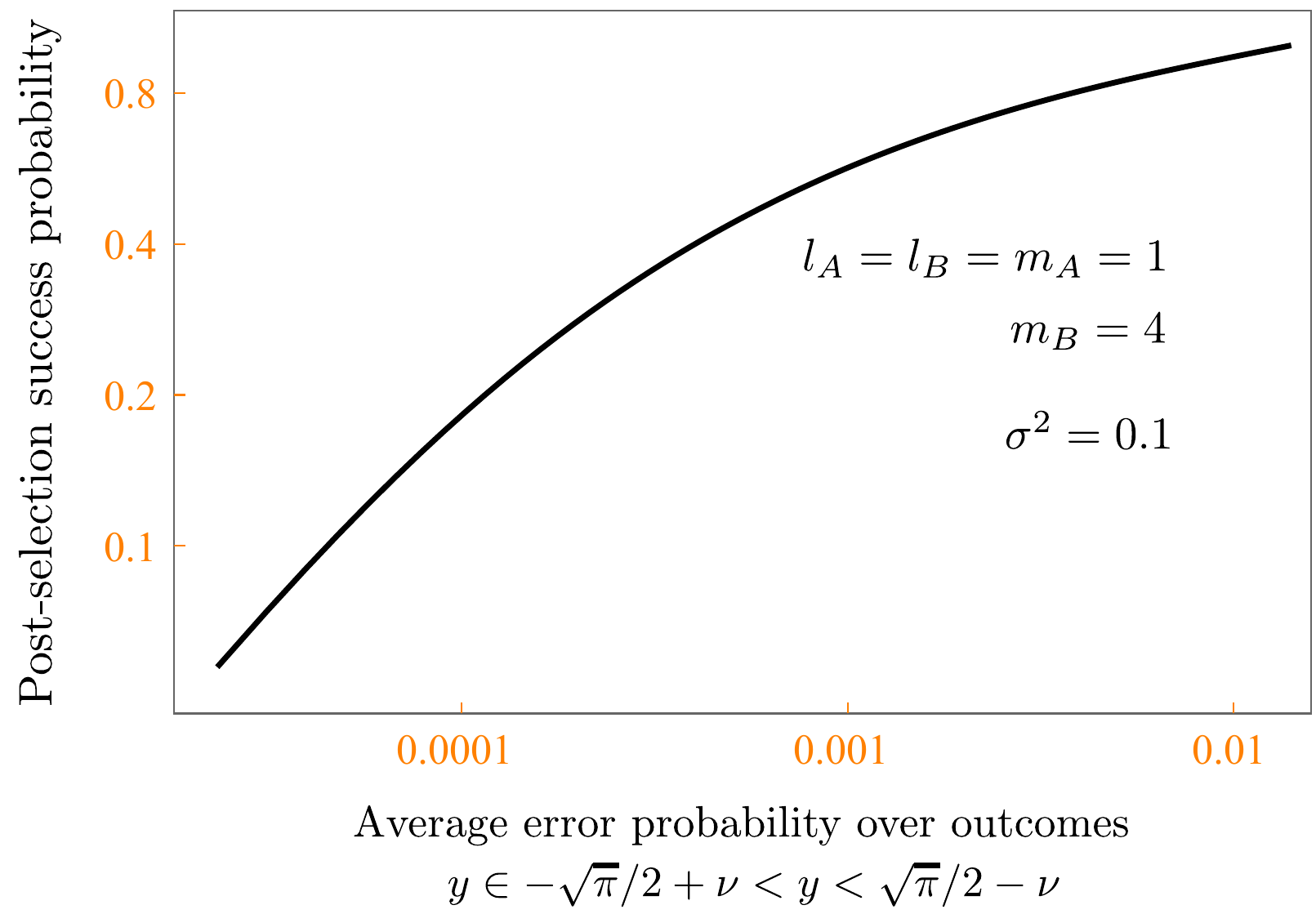}
    \caption{Post-selection success probability versus average error probability tradeoff by varying the post-selection exclusion window width $\nu$ in (\ref{psucc_vs_nu}), for  $l_A=l_B=m_A=1$, $m_B=4$ and $\sigma^2=0.1$ in (\ref{Steane_vertex_lm_values}).}
    \label{fig:psucc_vs_avgerror}
\end{figure}

Of course, this enhancement comes with an associated post-selection success probability, a function of $\nu$, given by
\begin{align}
P_{\textrm{succ}}(\nu)=\int_{I_0(\nu)} dyP_{Y}(y)+\int_{I_1(\nu)} dyP_{Y}(y),\label{psucc_vs_nu}
\end{align} 
where $I_0(\nu)\equiv|y|\leq \sqrt{\pi}/2-\nu$ and $I_1(\nu)\equiv\sqrt{\pi}>|y|\geq \sqrt{\pi}/2+\nu$.
Figure~\ref{fig:psucc_vs_nu} plots the postselection success probability versus $\nu$ in (\ref{psucc_vs_nu}) for  $l_A=l_B=m_A=1$, $m_B=4$ and $\sigma^2=0.1$ in (\ref{Steane_vertex_lm_values}), while Fig.~\ref{fig:psucc_vs_avgerror} plots the post-selection success probability versus average error probability tradeoff that ensues when varying $\nu$.

\subsection{Steane error correction on a vertex of a generic graph state}
Now, consider a general finite-energy approximate GKP qubit graph state of the form in (\ref{GKPgraphstate}) and $p-$quadrature Steane error correction of an arbitrary vertex $T$ of the graph using ancilla $A$. 
When the outcome of the measurement on the ancilla is $y$ such that $|y|<\sqrt{\pi}/2$ and when $m_A\ll m_T$, the Graph state transforms as $|\widetilde{\Psi_G(y)}\rangle$
\begin{align}
\approx\frac{\sqrt{P_N[0]P_Q(|y|)}|\widetilde{\Psi_0}\rangle+\sqrt{P_N[1]P_Q(\sqrt{\pi}-|y|)}|\widetilde{\Psi_1}\rangle}{\sqrt{P_N[0]P_Q(|y|)+P_N[1]P_Q(\sqrt{\pi}-|y|)}},\label{Steane_graphstate_output}
\end{align}
where
\begin{align}
|\widetilde{\Psi_0}\rangle&=\int d\vec{s}\,d\vec{t}\, \xi_G(\vec{s},\vec{t})\,e^{i(-\vec{s}.\vec{\hat{p}}+\vec{t}.\vec{\hat{q}})}|\overline{\Psi_G}\rangle,\\
|\widetilde{\Psi_1}\rangle&=\int d\vec{s}\,d\vec{t}\, \xi_G(\vec{s},\vec{t})\,e^{i(-(\vec{s}+\vec{s'}).\vec{\hat{p}}+(\vec{t}+\vec{t'}).\vec{\hat{q}})}|\overline{\Psi_G}\rangle\nonumber\\
&=\int d\vec{s}\,d\vec{t}\, \xi_G(\vec{s}-\vec{s'},\vec{t}-\vec{t'})\,e^{i(-\vec{s}.\vec{\hat{p}}+\vec{t}.\vec{\hat{q}})}|\overline{\Psi_G}\rangle
\end{align}
(up to phase factors $\exp(-is_Tl_A/(l_A+l_T)n\sqrt{\pi}),n\in\{0,1\}$) and $P_N[n]$ and $P_Q(p)$ are Gaussian distributions  of integer-valued and real-valued random variables $N$ and $Q$ given in (\ref{Steane_PN}) and (\ref{Steane_PQ}), respectively, with $\{l_B,m_B\}$ replaced by $\{l_T,m_T\}$. 
The covariance matrix and mean displacement elements of the error wavefunctions of $|\widetilde{\Psi_0}\rangle$ and $|\widetilde{\Psi_1}\rangle$ follow from standard Gaussian dynamics~\cite{GLS16}.
The vectors $\vec{s'},\vec{t'}$ in the latter are residual displacement errors on the qubits, whose entries are 0 except at $T$ (target), and for all first (i.e. nearest) neighboring vertices and second (i.e. next-to-nearest) neighboring vertices. 
%The $g$ values of Table~\ref{tab:Steane_fb_error} are in fact the gain factors used in applying the feedback displacement on the graph state qubits as part of the Steane error correction procedure. 
Thus, it is noteworthy that feedback displacements (to get rid of measurement outcome dependence on the mean displacements) only need to be performed up to the second nearest neighbors of the target.

Thus, under Steane error correction, a finite-energy GKP qubit graph state transforms into a conditional output, which is a superposition of finite-energy GKP qubit graph states whose error wavefunctions $\xi_G(\vec{x})$ are given by square roots of Gaussian distribution functions with identical covariance matrices of the form in (\ref{Graph_V_structure}), but with different mean displacement vectors. The state $|\widetilde{\Psi_1}\rangle$ represents the displacement error term in the state $|\widetilde{\Psi_G(y)}\rangle$ of (\ref{Steane_graphstate_output}) and the probability associated with this error is given by (\ref{eq:cn})-(\ref{Steane_logical_error_prob}), which is a function of the homodyne measurement outcome $y$. 
Note that post-selection can be used to reduce the error probability similarly as discussed for the single qubit case earlier. 

\begin{figure}
    \centering
    \includegraphics[width=0.45\textwidth]{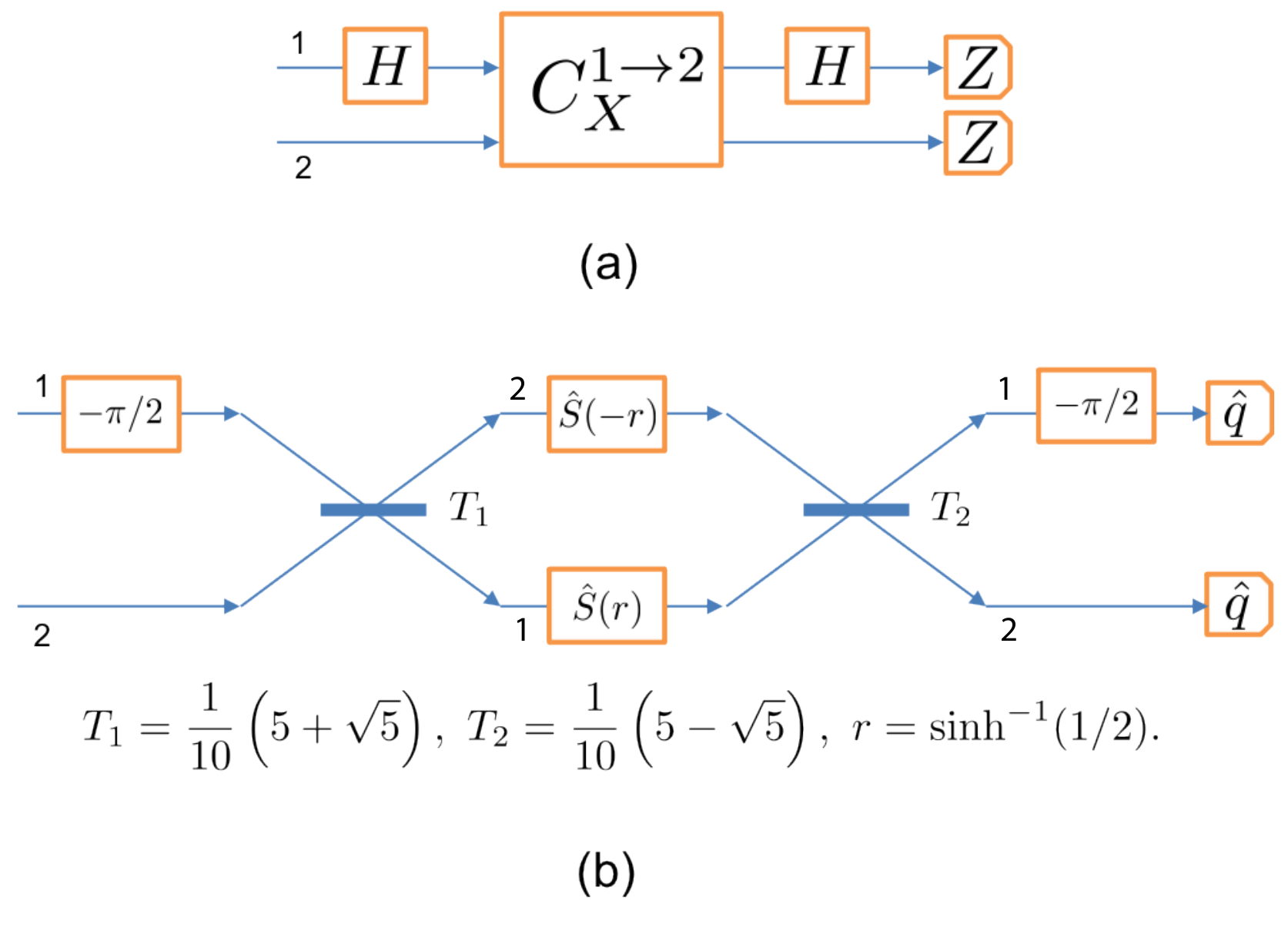}
    \caption{Fusion A: (a) circuit diagram and (b) its CV implementation for GKP qubits.}
    \label{fig:fusionA}
\end{figure}

\begin{figure}
    \centering
    \includegraphics[width=0.45\textwidth]{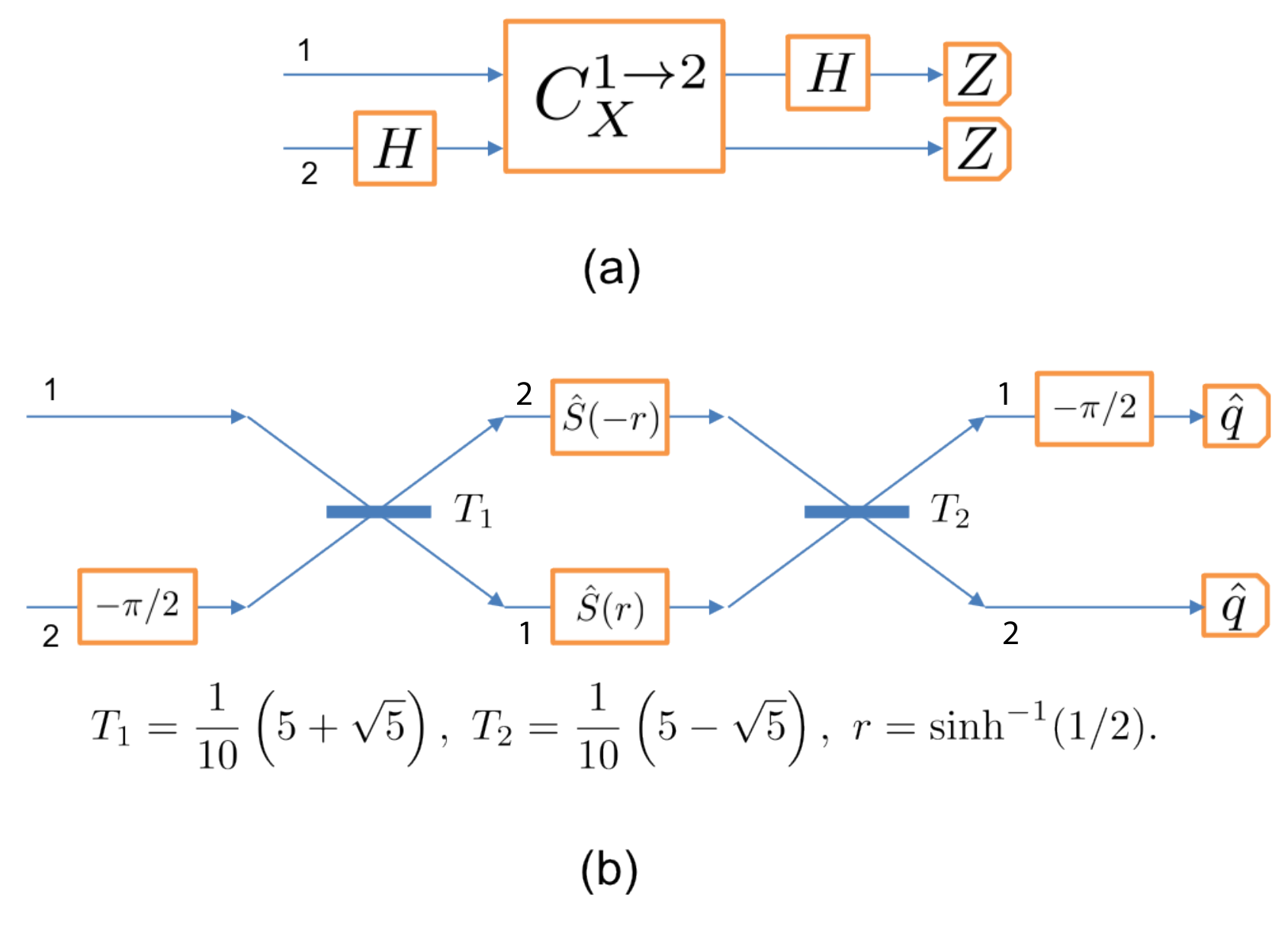}
    \caption{Fusion B: (a) circuit diagram and (b) its CV implementation for GKP qubits.}\label{fig:fusionB}
\end{figure}

\begin{figure}
    \centering
    \includegraphics[width=0.3\textwidth]{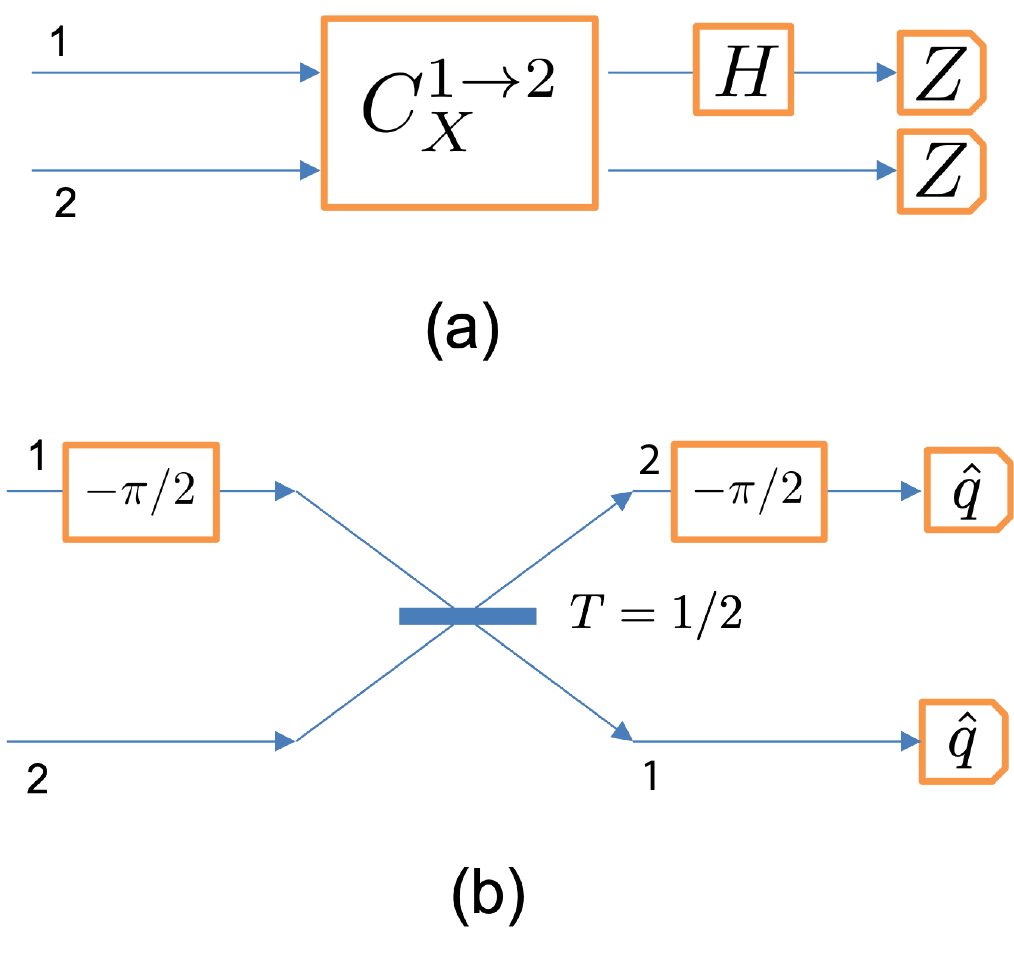}
    \caption{Fusion C: (a) circuit diagram and (b) its CV implementation for GKP qubits.}
    \label{fig:fusionC}
\end{figure}

\section{Fusion operations}	\label{fusion_ops}
In linear optical quantum computing (LOQC) with single-photon-based qubits, graph states that are universal for measurement-based quantum computing can be created by combining small graph states using what are referred to as fusion operations acting on photonic qubits~\cite{BR05, VBR07}. 
Here, we explore the CV analogues of the so-called "Type-II" fusion operations~\cite{BR05} from LOQC, which are rotated versions of maximally entangled two-qubit (Bell state) measurements, to apply on GKP qubits and fuse small GKP-qubit graph states to generate larger graph states of arbitrary topology~\footnote{Note that the other type of fusion operations from LOQC, namely Type-I, are a restricted version whose utility is limited to only growing linear graph states~\cite{BR05}. Since $C_Z$ gates can be used to produce small linear GKP qubit graph states, we will ignore Type-I fusion here.}. 
The fusion operations allow us to do so without rendering the GKP qubits in the graph state too noisy and prone to logical errors as would be the case if they were to be generated with $C_Z$ gates alone. 
Qubit circuits of three instances of Type-II fusion from LOQC, denoted as fusions A, B, and C, respectively, along with their CV analogues are shown in Figs.~\ref{fig:fusionA}, \ref{fig:fusionB} and \ref{fig:fusionC}. 
In these figures, $H$ denotes the Hadamard gate, $C_X$ denotes the controlled-NOT gate and $Z$ measurement denotes standard-basis measurement. 
The CV circuits involve the Fourier gate, beam splitters, squeezers, and $q-$quadrature homodyne detection for standard-basis measurement. 
The $-\pi/2$ gate denotes the Fourier gate, i.e., a rotation in phase space that transforms $\hat{q}\rightarrow \hat{p}$ and $\hat{p}\rightarrow -\hat{q}$. A beamsplitter of transmissivity $T$ transforms the input quadratures as
\begin{align}
			 \hat{q}_{1}&=\sqrt{T}\hat{q}_{1}+\sqrt{1-T}\hat{q}_{2}\nonumber\\
			\hat{p}_{1}&=\sqrt{T}\hat{p}_{1}+\sqrt{1-T}\hat{p}_{2}\nonumber\\
			\hat{q}_{2}&=-\sqrt{1-T}\hat{q}_{1}+\sqrt{T}\hat{q}_{2}\nonumber\\
			\hat{p}_{2}&=-\sqrt{1-T}\hat{p}_{1}+\sqrt{T}\hat{p}_{2}.
	\end{align}
The $\hat{S}(r)$ operation denotes a single-mode squeezer that transforms $\hat{q}\rightarrow e^{-r}\hat{q}$ and $\hat{p}\rightarrow e^{r}\hat{p}$. 
The equivalence of the CV circuits to the qubit circuits in the figures follow from the above the transformations of the input mode quadratures.
Note that while fusions A and B require inline squeezing, fusion C can be implemented without it. Fusion C implements what is known as dual homodyne measurement, which is used, e.g., in CV Gaussian entanglement swapping. 

An important distinction between single-photon based qubits and GKP qubits regarding the action of the fusion circuits is that whereas in the former the fusions fail when particular photon detection patterns are not observed at the detectors~\cite{VBR07}, the CV fusion operations on GKP qubits always succeed unless post-selection is performed on the measurement outcomes.
Post-selection on the measurement outcomes as part of the CV fusion operations could be performed in a manner identical to what was described earlier in Sec.~\ref{singlequbit QEC} A in the context of Steane error correction, to reduce logical errors in the resulting graph state qubits, but at the expense of rendering the fusion probabilistic. 
It is thus also possible to explore a tradeoff between the success probability of GKP fusion operations and the ensuing logical errors in the resulting graph state similar to the tradeoff presented for Steane error correction. 

When acted on ideal infinite-energy GKP qubits, all the 3 fusions mentioned above perform rotated Bell-state measurements given by the set of projectors $\{|\psi_{i,j}\rangle\langle\psi_{i,j}|,\ (i,j)\in\{0,1\}^2\},$ where
\begin{align}
    |\psi_{00}\rangle&=|0,+\rangle+|1,-\rangle=|+,0\rangle+|-,1\rangle,\nonumber\\
    |\psi_{10}\rangle&=|0,+\rangle-|1,-\rangle=|+,0\rangle-|-,1\rangle,\nonumber\\
     |\psi_{01}\rangle&=|0,-\rangle+|1,+\rangle=|-,0\rangle+|+,1\rangle,\nonumber\\
    |\psi_{11}\rangle&=|0,-\rangle-|1,+\rangle=|-,0\rangle-|+,1\rangle.\label{BSM_projs}
\end{align}
However, when acted on finite-energy approximate GKP qubits, the projections applied by these circuits are no longer exactly equivalent. 
They result in conditional output states with error wavefunctions having the same covariance matrix, but different mean displacements. 
We note that the fusions are to be followed by suitable feedback displacements on the vertices of the fused graph, which can be chosen in such a way as to remove the dependence of the resulting graph state on the measurement outcomes. Similar to Steane error correction, it turns out that feedback displacements are required to be performed only up to the second nearest neighbors of the vertices corresponding to the control qubit $C$ and the target qubit $T$ in the input sub-graph states.

When two finite-energy approximate GKP qubit graph states are fused using a fusion operation, the structure of the resulting graph state is governed by the action of the fusion on the underlying ideal GKP qubit graph state. As a result of their identical actions on ideal GKP qubits, all the three types of fusions mentioned above yield the same underlying graph structure. It is the same as the structure of the graph state resulting from the action of linear optical fusion operations on single-photon-based qubits~\cite{Ashlesha_paper}.

To further elucidate the action of the fusion operations, consider 2 graph states $|\widetilde{\Psi_{G_j}}\rangle=$ 
\begin{align}
 &\int d\vec{x}\ \eta_{G_j}(\vec{\mu},V,\vec{x})\prod_{i=1}^{n}e^{\frac{ix_ix_{n+i}}{2}}\hat{X}(x_i)\hat{Z}(x_{n+i})|\overline{\Psi_{G_j}}\rangle,\nonumber\\
 & j\in\{1,2\}
\end{align}
of the form in (\ref{GKPgraphstate}), as described in Section~\ref{GKPgraph}, and identify two vertices $C$ and $T$, one from each of the two sub-graphs. 
When a fusion operation is applied from $C$ to $T$, and the measurement outcomes are $|y_C|, \ |y_T|\leq \sqrt{\pi}/2$, the post-fusion (followed by feedback displacements on up to second-nearest neighboring vertices of $C$ and $T$), the state is given by $|\widetilde{\Psi'_{G}}\rangle|_{y_C,y_T}\propto\approx$
\begin{align}
& \sqrt{P_{N_C}[0]P_{Q_C}(|y_C|)P_{N_T}[0]P_{Q_T}(|y_T|)}|\widetilde{\psi_{G}^{'(00)}}\rangle\nonumber\\
&+ \sqrt{P_{N_C}[0]P_{Q_C}(|y_C|)P_{N_T}[1]P_{Q_T}(\sqrt{\pi}-|y_T|)}|\widetilde{\psi_{G}^{'(01)}}\rangle\nonumber\\
&+\sqrt{P_{N_C}[1]P_{Q_C}(\sqrt{\pi}-|y_C|)P_{N_T}[0]P_{Q_T}(|y_T|)}|\widetilde{\psi_{G}^{'(10)}}\rangle\nonumber\\
&+\sqrt{P_{N_C}[1]P_{Q_C}(\sqrt{\pi}-|y_C|)P_{N_T}[1]P_{Q_T}(\sqrt{\pi}-|y_T|)}\nonumber\\
&\times|\widetilde{\psi_{G}^{'(11)}}\rangle\label{fusion_output_superposition}
\end{align}
where $|\widetilde{\psi_{G}^{'(uv)}}\rangle\equiv G(\mathrm{V'},\mathrm{E'},\vec{\mu}^{'(uv)},V'),\ u,v\in\{0,1\}$, $|\mathcal{V'}|=n-2$ and the topology of the fused graph $\mathrm{E'}$ can be found in Ref.~\cite{Ashlesha_paper}.
%The homodyne outcome probability distributions $P_{N_C}[n\sqrt{\pi}],P_{N_T}[n\sqrt{\pi}],P_{Q_C}(p),P_{Q_T}(p)$ and 
The transformation of the error wave function covariance matrix and mean displacement elements of these conditional graph states $|\widetilde{\psi_{G}^{'(uv)}}\rangle$, namely $V',\vec{\mu}^{'(uv)}$, under the fusion operations A, B and C in terms of the pre-fusion covariance matrix elements and mean displacement vector follow from standard Gaussian dynamics~\cite{GLS16}. All terms in the superposition in (\ref{fusion_output_superposition}) except the one corresponding to $u=v=0$ are error terms, whose probabilities are given by their coefficients in the superposition upto suitable normalization. The error probabilities are functions of $y_C, y_T$, and similar to Steane error correction, could be reduced using post selection.

\begin{figure}
    \centering
    \includegraphics[width=0.45\textwidth]{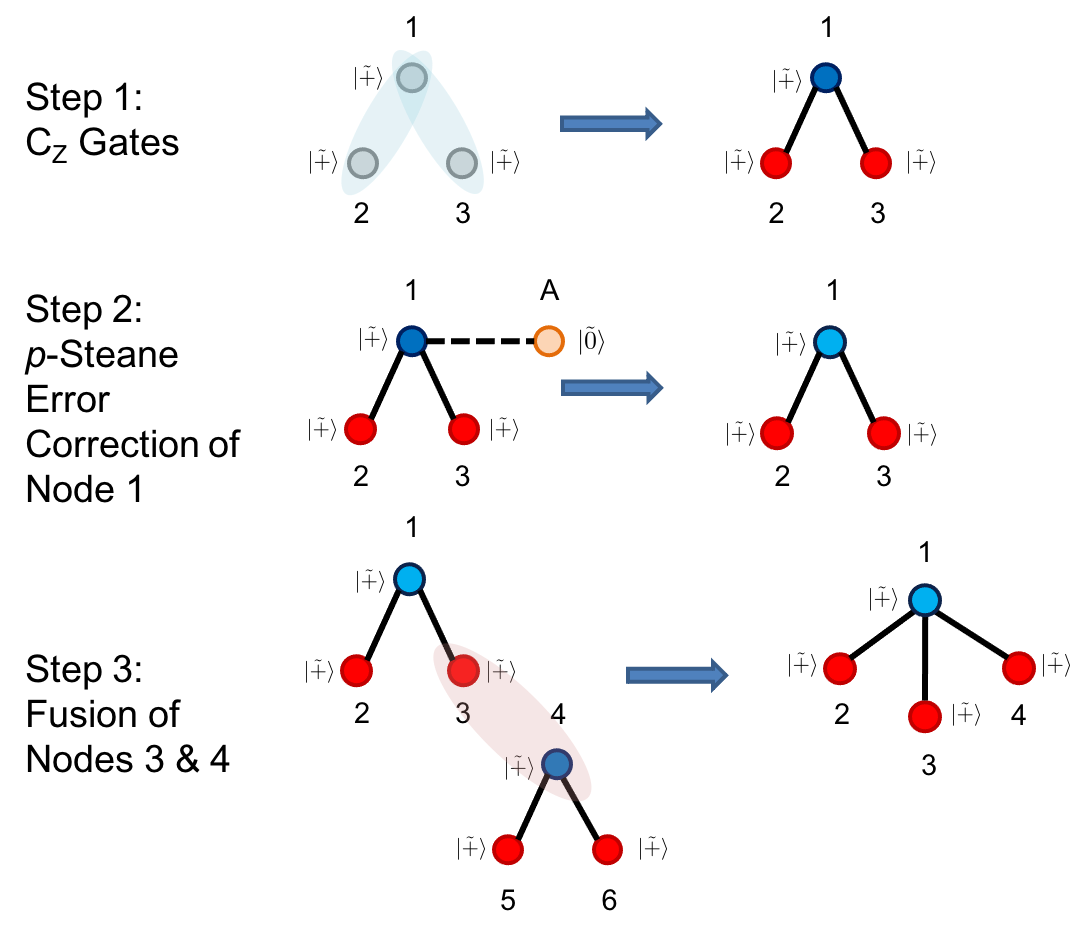}
    \caption{Generation of a 4-qubit tree graph state from finite-energy approximate GKP qubits using $C_Z$ gate, Steane error correction and Fusion operations, as per the protocol presented Ref.~\cite{FTOF18}}.
    \label{fig:Fukui_example}
\end{figure}
\begin{table}[]
    \centering
    \begin{tabular}{|P{0.87cm}|P{7.0cm}|}
        %\hline
       % \multicolumn{2}{|c|}{Error wavefunction Covariance Matrix for different Fusions} \\
        \hline
        Fusion & Covariance Matrix\\
%         \hline
%         A & 
% $\begin{pmatrix}
%  \frac{3 }{2} & 0 & 0 & 0 & 0 & -\frac{1}{2} & 0 & 0 \tabularnewline
%  0 & \frac{3 }{4} & 0 & 0 & -\frac{1}{4} & 0 & 0 & 0\tabularnewline
%  0 & 0 & \frac{2 }{3} & -\frac{1}{3} & 0 & 0 & 0 & 0\tabularnewline
%  0 & 0 & -\frac{1}{3} & \frac{2 }{3} & 0 & 0 & 0 & 0\tabularnewline
%  0 & -\frac{1}{4} & 0 & 0 & \frac{3 }{4} & 0 & 0 & 0\tabularnewline
%  -\frac{1}{2} & 0 & 0 & 0 & 0 & \frac{3 }{2} & 0 & 0\tabularnewline
%  0 & 0 & 0 & 0 & 0 & 0 & 2  & 1 \tabularnewline
%  0 & 0 & 0 & 0 & 0 & 0 & 1 & 2
% \end{pmatrix}\sigma ^2$
% \\
%         \hline
%         B & 
% $\begin{pmatrix}
%  2 & 0 & 0 & 0 & 0 & -1 & 0 & 0 \tabularnewline
%  0 & \frac{2}{3} & 0 & 0 & -\frac{1}{3} & 0 & 0 & 0 \tabularnewline
%  0 & 0 & 1 & 0 & 0 & 0 & 0 & 0 \tabularnewline
%  0 & 0 & 0 & 1 & 0 & 0 & 0 & 0 \tabularnewline
%  0 & -\frac{1}{3} & 0 & 0 & \frac{2}{3} & 0 & 0 & 0 \tabularnewline
%  -1 & 0 & 0 & 0 & 0 & 2  & 0 & 0 \tabularnewline
%  0 & 0 & 0 & 0 & 0 & 0 & 1 & 0 \tabularnewline
%  0 & 0 & 0 & 0 & 0 & 0 & 0 & 1
% \end{pmatrix}\sigma ^2$
% \\
        \hline
        A,B,C & 
$\begin{pmatrix}
 \frac{5}{3} & 0 & 0 & 0 & 0 & -\frac{2}{3} & \frac{1}{3} & \frac{1}{3} \tabularnewline
 0 & \frac{11}{15} & \frac{1}{15} & \frac{1}{15} & -\frac{4}{15} & 0 & 0 & 0 \tabularnewline
 0 & \frac{1}{15} & \frac{11 }{15} & -\frac{4}{15} & \frac{1}{15} & 0 & 0 & 0 \tabularnewline
 0 & \frac{1}{15} & -\frac{4}{15} & \frac{11}{15} & \frac{1}{15} & 0 & 0 & 0 \tabularnewline
 0 & -\frac{4}{15} & \frac{1}{15} & \frac{1}{15} & \frac{11}{15} & 0 & 0 & 0 \tabularnewline
 -\frac{2}{3}  & 0 & 0 & 0 & 0 & \frac{5}{3} & -\frac{1}{3} & -\frac{1}{3} \tabularnewline
 \frac{1}{3} & 0 & 0 & 0 & 0 & -\frac{1}{3} & \frac{5}{3} & \frac{2}{3} \tabularnewline
 \frac{1}{3} & 0 & 0 & 0 & 0 & -\frac{1}{3} & \frac{2}{3} & \frac{5}{3} 
\end{pmatrix}\sigma ^2$
\\
        \hline
    \end{tabular}
    \caption{Covariance Matrix of the error wavefuctions at the output of the protocol described in Fig.~\ref{fig:Fukui_example}. It is the same independent of the fusion applied. The quadratures are ordered as $(q_1,\ldots q_4,p_1,\ldots p_4)$.}
    \label{tab:Fukui_protocol_covariances}
\end{table}

\begin{table}[]
    \centering
    \begin{tabular}{|P{0.875cm}|P{7.0cm}|}
        %\hline
       % \multicolumn{2}{|c|}{Error wavefunction Covariance Matrix for different Fusions} \\
        \hline
        Fusion & Mean Displacement Vectors\\
        \hline
        A & 
$\big\{-\frac{\sqrt{\pi }}{3} u,\frac{\sqrt{\pi }}{15}(-4w+v),\frac{\sqrt{\pi }}{15}(w-4v),\frac{\sqrt{\pi }}{15}(w-4v),$\\
&$\frac{\sqrt{\pi }}{15}(11w+v),\frac{\sqrt{\pi } u}{3},\frac{\sqrt{\pi } u}{3},\frac{ \sqrt{\pi } u}{3}\big\}$
\\
        \hline
        B & 
$\big\{\frac{\sqrt{\pi }}{3}v,\frac{\sqrt{\pi }}{15}(-4w+u),\frac{\sqrt{\pi }}{15}(w-4u),\frac{\sqrt{\pi }}{15}(w-4u),$\\
&$\frac{\sqrt{\pi }}{15}(11w+u),-\frac{\sqrt{\pi }}{3}v,-\frac{\sqrt{\pi }}{3}v,-\frac{\sqrt{\pi }}{3}v\big\}$
\\
        \hline
        C & 
$\big\{\frac{\sqrt{2 \pi }}{3}  u,\frac{\sqrt{\pi }}{15}  \left(-4 w+\sqrt{2} v\right),-\frac{\sqrt{\pi }}{15}  \left(w-4 \sqrt{2} v\right),$\\
&
$-\frac{\sqrt{\pi }}{15}  \left(w-4 \sqrt{2} v\right),\frac{\sqrt{\pi }}{15}  \left(11 w+\sqrt{2} v\right),$\\
&$
-\frac{\sqrt{2 \pi }}{3}  u,\frac{\sqrt{2 \pi }}{3}  u,\frac{\sqrt{2 \pi }}{3}  u\big\}$
\\
        \hline
    \end{tabular}
    \caption{Error wavefuction mean displacement vectors of the finie-energy GKP-qubit graph states that are in coherent superposition at the output of the protocol described in Fig.~\ref{fig:Fukui_example}, for the different choice of fusion operations A, B or C. The quadratures are ordered as $(q_1,\ldots q_4,p_1,\ldots p_4)$. The different tuples of indices $u,v,w\in\{0,1\}$ correspond to the 8 terms that are in superposition.}
    \label{tab:Fukui_protocol_meanvectors}
\end{table}

%We will now analyze how the different types of fusion operations ``fuse" these sub-graph states into one.

%\subsubsection*{Type A fusion}
%The type A fusion implements the 2-qubit projection given by $\frac{\bra{+0}\pm\bra{-1}}{\sqrt{2}}$. The corresponding discrete variable qubit circuit and optical circuit are shown in Fig.~\ref{fig:fusionA}. The optical circuit implements the said projection exactly when applied on two infinitely squeezed ideal GKP qubits. When performed on two cluster states, fusion A connects all neighbours of qubit $C$ and qubit $T$ with each other~\cite{Ashlesha_paper}. The transformation rules for the error wave function covariance matrix and mean displacement elements under fusion A are tabulated in Table~\ref{}. %\Ashlesha{(cite paper in preparation)}.

\begin{figure}
    \centering
    \includegraphics[width=0.45\textwidth]{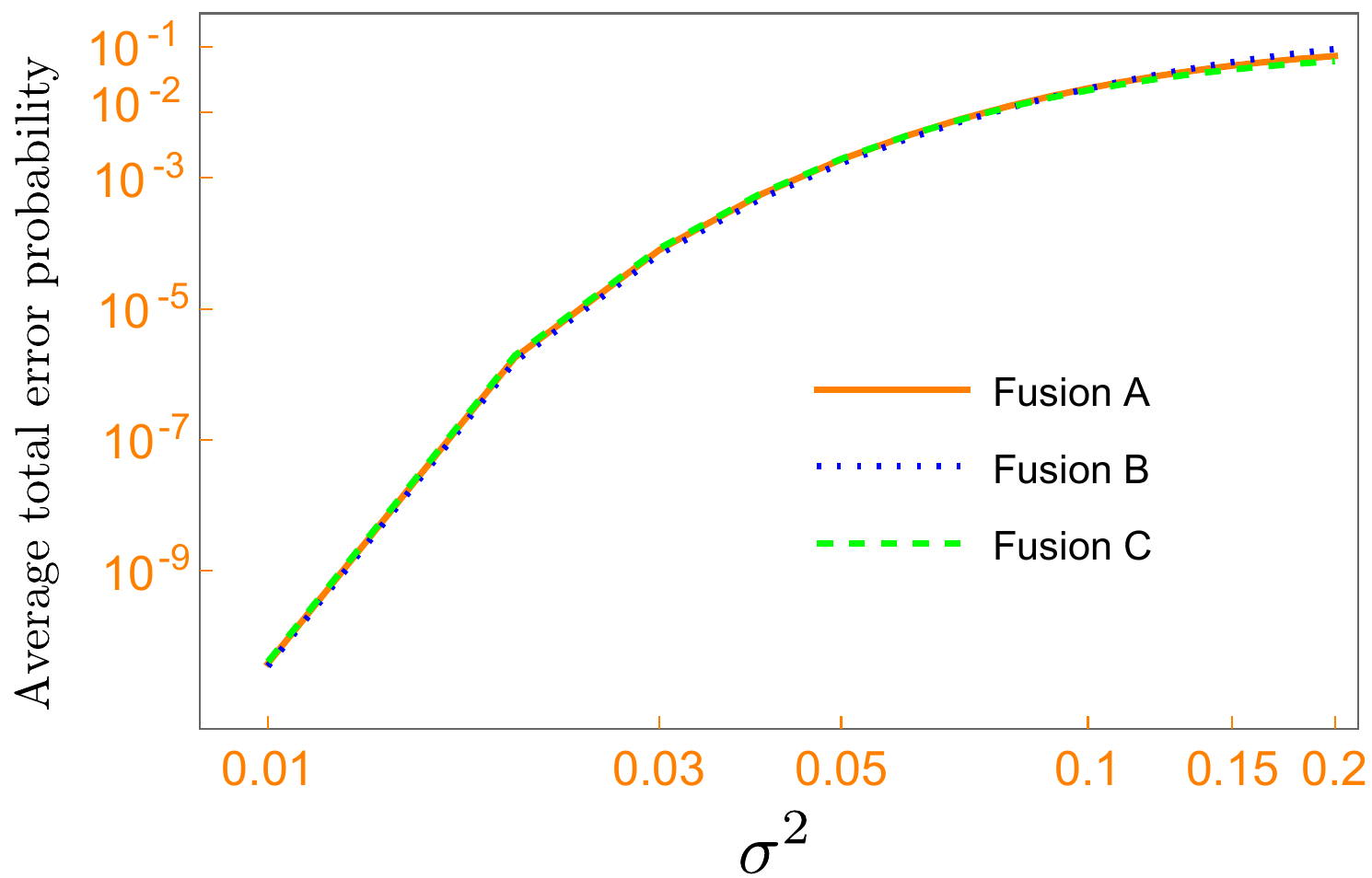}
    \caption{Average total error probability associated with the generation of the 4-qubit tree graph state from finite-energy approximate GKP qubits using the protocol presented Ref.~\cite{FTOF18} with the three different fusions discussed in Sec~\ref{fusion_ops}, as a function of $\sigma^2$, where the initial finite-energy approximate GKP qubits have teeth and envelope variances given by $\sigma^2/2,1/2\sigma^2$, respectively.}
    \label{fig:Fukui_ex_avg_errorprob}
\end{figure}

\section{Discussion}\label{discussions}
The tools discussed in Sections~\ref{singlequbit QEC} and \ref{fusion_ops} can be used to generate fault-tolerant graph states starting from finite-energy approximate GKP qubit $|\widetilde{+}\rangle$ states. 
Ref.~\cite{FTOF18} provided a protocol to generate graph states starting from mixed state GKP qubits that are defined as incoherent Gaussian mixtures of randomly displaced ideal GKP qubit states. 
Such states can be obtained from the pure finite-energy GKP qubit states considered in this work by a Gaussian displacement twirling operation, and thus by the data-processing inequality, are more noisy. 
An approach similar to the one in Ref.~\cite{FTOF18} can be adopted to generate universal GKP graph states from the pure, finite energy approximate GKP qubit states considered here using fusions and Steane error correction in a ballistic fashion similar to discrete-variable linear optical schemes~\cite{PKES17}.

%calculation for Steane error correction on a 5-qubit star cluster state followed Type-C fusion between the leaf qubits of two copies of this state in Table~\ref{table:4star} and Table~\ref{table:fuse4star}, respectively. Our work can also be used to design all-optical quantum repeaters (cite Mihir) using GKP qubits.

As a demonstration of our analysis, we look at the generation of a small 4-qubit tree cluster made of finite-energy approximate GKP qubits, tracing the first few steps of the protocol followed in Ref.~\cite{FTOF18}. 
The protocol is described step by step in Fig.~\ref{fig:Fukui_example}. 
While the analysis in Ref.~\cite{FTOF18} tracked the individual quadrature noise variances of mixed state GKP qubits, our analysis with truly finite energy GKP qubits tracks the full covariance matrix of the Gaussian error wavefunction of the graph state along with the mean displacement vector. This work thus provides a more accurate analysis of the  errors that are introduced during the graph creation from approximate GKP qubit pure states due to (a) the finite-energy approximation, and (b) 
%the misidentification of the underlying ideal GKP qubit states from homodyne detection outcomes.
homodyne measurements that are part of the graph state generation protocol. Since the 4-qubit tree cluster generation as per the protocol involves 1 steane error correction and 1 fusion operation, there are 3 homodyne measurements. This results in a total of $8$ terms in superposition at the ouptut, which correspond to finite-energy GKP qubit graph states whose error wavefunctions $\xi_G(\vec{x})$ are given by square roots of Gaussian distribution functions with identical covariance matrices of the form in (\ref{Graph_V_structure}) given in Table~\ref{tab:Fukui_protocol_covariances}, but with different mean displacement vectors, as tabulated in Table~\ref{tab:Fukui_protocol_meanvectors} indexed by $u,v,w\in\{0,1\}$. The total error probability, i.e., the norm of the weights associated with all but the term corresponding to $u=v=w=0$ in the superposition, averaged over the outcomes homodyne measurement outcomes in the Steane error correction and the fusion, are plotted for the 3 fusions in Fig.~\ref{fig:Fukui_ex_avg_errorprob} as function of initial GKP-qubit squeezing variance $\sigma^2$. 
We observe that though the output states from the three fusions are not exactly identical, the error probabilities are. 
The error probabilities can be further reduced by considering post-selected homodyne measurements as part of both the Steane error correction and fusion, as discussed earlier in Sec.~\ref{singlequbit QEC} A.

In summary, we presented an exact description of graph states composed of truly finite-energy, approximate GKP qubit pure states in terms of Gaussian error wavefunctions. We tracked the transformation of the error wavefunction's covariance matrix and mean vector under Steane error correction and graph fusion operations that are used to generate high-fidelity large graph states. The output of these procedures in the coherent error wavefunction description are coherent superpositions of $2^n$ number of approximate GKP qubit graph states ($n$ being the number of homodyne measurement involved) whose error wavefunctions have identical covariance matrices, but different mean displacement vectors, all of which except one correspond to mean displacement errors in phase space. The error probabilities are functions of the homodyne measurement output statistics and can be reduced using post-selection generation of graph states at the expense of finite success probability of graph state generation. Whereas studies hitherto on GKP qubit graph states have dealt with incoherent, mixed state descriptions of GKP qubits, our work presents an accurate model for the noise and displacement errors present in graph states composed of finite-energy GKP qubits. Our work thus could potentially be useful in generating and characterizing the error correction properties of large graph states for measurement-based quantum information processing with applications in quantum computing and all-optical quantum repeaters.

\section*{Acknowledgments}
K.P.S. and P.D. acknowledge funding support from a Department of Energy (DoE) project on continuous variable quantum repeaters, funded under a subcontract from ORNL, subcontract number 4000178321. A.P., L.J. and S.G. acknowledge support of the National Science Foundation (NSF) ERC, Center for Quantum Networks, award number 1941583. L.J. acknowledges support from the ARO (W911NF-18-1-0020, W911NF-18-1-0212), ARO MURI (W911NF-16-1-0349, W911NF-16-1-0349, W911NF-21-1-0325), AFOSR MURI (FA9550-19-1-0399), NSF (EFMA-1640959, OMA-1936118, OMA-2137642, EEC-1941583), NTT Research, and the Packard Foundation (2013-39273). S.G., K.P.S. and A.P. gratefully acknowledge several useful discussions with Rafael Alexander, and various researchers at Xanadu Quantum Technologies, esp., Ish Dhand, Krishnakumar Sabapathy, Guillaume Dauphinais, and Ilan Tzitrin. K.P.S. gratefully acknowledges several useful discussions with Filip Rozpedek. 
\section*{Conflict of Interest}

S.G. serves as a scientific advisor for Xanadu Quantum Technologies, a company pursuing fault-tolerant quantum computing with photonic GKP qubits. S.G. owns stock options in the company and has received financial compensation for his technical advisory services. The other authors declare no conflicts of interest.

\appendix

\section{Equivalence between error wavefunction and quadrature-basis descriptions of GKP qubit states.\label{GKP_equivalence}}
Here we show the equivalence between the error wavefunction description and quadrature-basis description of a GKP qubit state. Consider as an example, the $|\widetilde{+}\rangle$ state. Without loss of generality, assuming zero mean displacements, we have
\begin{align}
|\widetilde{+}\rangle&=\int du dv \frac{1}{\sqrt{\pi \kappa\delta}}\exp(-\frac{1}{2}(u^2/\delta^2+v^2/\kappa^2))\nonumber\\
&\exp(i(-u \hat{p} + v \hat{q}))|\overline{+}\rangle,\\
&=\int du dv \frac{1}{\sqrt{\pi \kappa\delta}}\exp(-\frac{1}{2}(u^2/\delta^2+v^2/\kappa^2))\nonumber\\
&\exp(iuv/2) \hat{X}(u)\hat{Z}(v)|\overline{+}\rangle,
\end{align}
which follows from (\ref{XZ_ZX_displ}). Proceeding further, we have
\begin{align}
|\widetilde{+}\rangle &=\sum_{n=-\infty}^\infty\frac{1}{\sqrt{\pi \kappa\delta}}\int du e^{\frac{iuv}{2}} e^{-\frac{u^2}{2\delta^2}}e^{-iu(2n\sqrt{\pi}+v)}\nonumber\\
&\times \int dv e^{-\frac{v^2}{2\kappa^2}}|p=2n\sqrt{\pi}+v\rangle,
\end{align}
where $|\overline{+}\rangle$ has been expanded in the basis of $p-$eigenstates, and the action of the $\hat{X}$ operator on these states results in  the factor $e^{-iu(2n\sqrt{\pi}+v)}$. The above state can be reexpressed as
\begin{align}
|\widetilde{+}\rangle&=\sum_{n=-\infty}^\infty\frac{\sqrt{2\pi\delta^2}}{\sqrt{\pi \kappa\delta}}\int du \frac{e^{-\frac{u^2}{2\delta^2}}}{\sqrt{2\pi\delta^2}}e^{-iu(2n\sqrt{\pi}-v/2)}\nonumber\\
&\int dv e^{-\frac{v^2}{2\kappa^2}}\hat{Z}(2n\sqrt{\pi}+v)|p=0\rangle\\
&=\sqrt{\frac{2\delta}{\kappa}}\sum_n\int dv e^{-\frac{(2n\sqrt{\pi}+v/2)^2\delta^2}{2}}e^{-\frac{v^2}{2\kappa^2}}\nonumber\\
&\hat{Z}(2n\sqrt{\pi}+v)|p=0\rangle,
\end{align}
which follows from evaluating the Fourier intergral in variable $u$. The above equation can be re-expressed with completion of squares in variable $v$ as
\begin{align}
|\widetilde{+}\rangle&=\sqrt{\frac{2\delta}{\kappa}}\sum_n e^{-\frac{8\pi\delta^2}{4+\kappa^2\delta^2}n^2}\int dv e^{-\frac{\left(v+\frac{4\sqrt{\pi}\delta^2\kappa^2n}{4+\delta^2\kappa^2}\right)^2}{\frac{8\kappa^2}{4+\delta^2\kappa^2}}}\nonumber\\
&\hat{Z}(2n\sqrt{\pi}+v)|p=0\rangle\\
&=\sqrt{2\pi}\sum_n \frac{e^{-\frac{(2\sqrt{\pi}n)^2}{2\left(\frac{4+\delta^2\kappa^2}{4\delta^2}\right)}}}{\left(\pi\left(\frac{4+\delta^2\kappa^2}{4\delta^2}\right)\right)^{1/4}}\int dv \frac{e^{-\frac{\left(v+\frac{4\sqrt{\pi}\delta^2\kappa^2n}{4+\delta^2\kappa^2}\right)^2}{\frac{8\kappa^2}{4+\delta^2\kappa^2}}}}{\left(\pi\left(\frac{4\kappa^2}{4+\delta^2\kappa^2}\right)\right)^{1/4}}\nonumber\\
&\hat{Z}(2n\sqrt{\pi}+v)|p=0\rangle.
\end{align}
In the limit $\kappa^2\delta^2\rightarrow 0$, we get
\begin{align}
|\widetilde{+}\rangle&=\sqrt{2\pi}\sum_n \frac{e^{-\frac{\delta^2(2\sqrt{\pi} n)^2}{2}}}{(\pi/\delta^2)^{1/4}}\int dv \frac{e^{-\frac{v^2}{2\kappa^2}}}{(\pi\kappa^2)^{1/4}}\nonumber\\
&\hat{Z}(2n\sqrt{\pi}+v)|p=0\rangle,
\end{align}
thus showing the equivalence between the error wavefunction and quadrature wavefunction descriptions.

%\section{Appendix: GKP graph states}
%This is because, we have
%\begin{align}
    %|\widetilde{+}\rangle&=\int dsdt \frac{e^{-\frac{1}{2}\left(\frac{(s-\bar{q})^2}{l}+\frac{(t-\bar{p})^2}{m}\right)}}{(\pi lm)^{1/4}} e^{i\frac{st}{2}}\hat{X}(s)\hat{Z}(t)|\overline{+}\rangle,\\
   % \otimes_{i=1}^{n}|\widetilde{+}\rangle_i &=\int d\vec{x} \frac{e^{-\frac{1}{2}(\vec{x}-\vec{\mu}).V^{-1}.(\vec{x}-\vec{\mu})^T}}{(\pi^n \det{V})^{1/4}} e^{i\frac{\vec{s}.\vec{t}}{2}}\hat{X}(\vec{s})\hat{Z}(\vec{t})\otimes_{i=1}^n|\overline{+}\rangle_i,\\
   % \textrm{where\ } \vec{x}&=(\vec{s},\vec{t}),\ \vec{s}=(s_1,\ldots,s_n),\vec{t}=(t_1,\ldots,t_n),\ \vec{\mu}=(\vec{\mu}_q,\vec{\mu}_p),\ \vec{\mu}_q=(\bar{q}_1,\ldots,\bar{q}_n),\vec{\mu}_p=(\bar{p}_1_1,\ldots,\bar{p}_n),\\
  %  V&=\operatorname{diag}(\vec{l},\vec{m}), \ \vec{l}=(l_1,\ldots,l_n),\ \vec{m}=(m_1,\ldots,m_n)\\
   % \Rightarrow C_Z^{G(n,A)}\otimes_{i=1}^{n}|\widetilde{+}\rangle_i &=\int d\vec{x} \frac{e^{-\frac{1}{2}(\vec{x}-\vec{\mu}).V^{-1}.(\vec{x}-\vec{\mu})^T}}{(\pi^n \det{V})^{1/4}} e^{i\frac{\vec{s}.\vec{t}}{2}}C_Z^{G(n,A)}\hat{X}(\vec{s})\hat{Z}(\vec{t}){C_Z^{G(n,A)}}^\dagger C_Z^{G(n,A)}\otimes_{i=1}^n|\overline{+}\rangle_i\\
  %  &=|\widetilde{\Psi_G}\rangle
%\end{align}
\section{Steane Error Correction Details \label{Steane_EC_details}}
When mode $A$ is measured over its $p-$quadrature, we get an outcome $y\in\mathbb{R}$ with probability $P_{Y}(y)$ and a conditional state $|\phi\rangle_{B|y}$ on mode B, that can be deduced from the conditional unnormalized state given by
\begin{align}
	&_p\langle y|_A.|\psi\rangle_{AB}=\sqrt{P_{Y}(y)}|\phi\rangle_{B|y}\nonumber\\
	&=\int ds_Ads_B \chi_{AB}(s_A,s_B)\int dt_Adt_B\mathcal{P}(t_A)\mathcal{Q}(t_B|t_A)\nonumber\\
	&_p\langle y|_A. e^{i(-s_A\hat{p}_A+t_A\hat{q}_A)}|\overline{0}\rangle_A e^{i(-s_B\hat{p}_B+t_B\hat{q}_B)}|\overline{+}\rangle_B\\
	&=\sum_n\int ds_Ads_B \chi_{AB}(s_A,s_B)\int dt_A dt_B\mathcal{P}(t_A)\mathcal{Q}(t_B|t_A)\nonumber\\
	&_p\langle y|_A. e^{\frac{is_At_A}{2}}\hat{X}_A(s_A)\hat{Z}_A(t_A){|n\sqrt{\pi}\rangle_p}_A e^{i(-s_B\hat{p}_B+t_B\hat{q}_B)}|\overline{+}\rangle_B\\	
	&=\sum_n\int ds_Ads_B \chi_{AB}(s_A,s_B)\int dt_Adt_B\mathcal{P}(t_A)\mathcal{Q}(t_B|t_A)\nonumber\\
	& e^{\frac{is_At_A}{2}} {_p}\langle y|_A.|\hat{X}_A(s_A){|n\sqrt{\pi}+t_A\rangle_p}_A e^{i(-s_B\hat{p}_B+t_B\hat{q}_B)}|\overline{+}\rangle_B\\
	&=\sum_n\int ds_Ads_B \chi_{AB}(s_A,s_B)\int dt_Adt_B\mathcal{P}(t_A)\mathcal{Q}(t_B|t_A)\nonumber\\
	& e^{-is_A(n\sqrt{\pi}+\frac{t_A}{2})}\delta(y-(n\sqrt{\pi}+t_A))e^{i(-s_B\hat{p}_B+t_B\hat{q}_B)}|\overline{+}\rangle_B\\
	&=\sum_n\int ds_Ads_B \chi_{AB}(s_A,s_B)\int dt_Adt_B\mathcal{P}(t_A)\mathcal{Q}(t_B|t_A)\nonumber\\
	& e^{-is_A(n\sqrt{\pi}+\frac{t_A}{2})}\delta(t_A-(y-n\sqrt{\pi}))e^{i(-s_B\hat{p}_B+t_B\hat{q}_B)}|\overline{+}\rangle_B\\
	&=\sum_n\int ds_Ads_Bdt_B \chi_{AB}(s_A,s_B)\mathcal{P}(t_A=y-n\sqrt{\pi})\nonumber\\
	&\mathcal{Q}(t_B|t_A=y-n\sqrt{\pi})e^{\frac{-is_A(y+n\sqrt{\pi})}{2}}e^{i(-s_B\hat{p}_B+t_B\hat{q}_B)}|\overline{+}\rangle_B.
	\end{align}
	Using the distributions for $\chi, \mathcal{P},\mathcal{Q}$ from (\ref{Steane_distributions}), we have $\sqrt{P_{Y}(y)}|\phi\rangle_{B|y}$
	\begin{align}
	&=\sum_n\frac{e^{-\frac{(y-n\sqrt{\pi})^2}{2(m_A+m_B)\sigma^2}}}{(\pi(m_A+m_B)\sigma^2)^{1/4}}\int ds_Bdt_B\int ds_A\chi_{AB}(s_A,s_B)\nonumber\\
	&e^{\frac{-is_A(y+n\sqrt{\pi})}{2}}\frac{e^{-\frac{\left(t_B+\frac{m_B}{m_A+m_B}(y-n\sqrt{\pi})\right)^2}{2\frac{m_Am_B}{m_A+m_B}\sigma^2}}}{(\pi\frac{m_Am_B}{m_A+m_B}\sigma^2)^{1/4}}e^{i(-s_B\hat{p}_B+t_B\hat{q}_B)}|\overline{+}\rangle_B
	\end{align}
	\begin{align}
	&=2\sqrt{\pi}\sum_n\frac{e^{-\frac{(y-n\sqrt{\pi})^2}{2(m_A+m_B)\sigma^2}}}{(\pi(m_A+m_B)\sigma^2)^{1/4}}\frac{e^{-\frac{(y+n\sqrt{\pi})^2}{8\frac{(l_A+l_B)}{l_Al_B\sigma^2}}}}{(4\pi\frac{(l_A+l_B)}{l_Al_B\sigma^2})^{1/4}}\nonumber\\
	&\times\Bigg(\int ds_Bdt_B \frac{e^{-\frac{s_B^2}{2(l_A+l_B)\sigma^2}}}{(\pi(l_A+l_B)\sigma^2)^{1/4}}\frac{e^{-\frac{\left(t_B+\frac{m_B}{m_A+m_B}(y-n\sqrt{\pi})\right)^2}{2\frac{m_Am_B}{m_A+m_B}\sigma^2}}}{(\pi\frac{m_Am_B}{m_A+m_B}\sigma^2)^{1/4}}\nonumber\\
	&e^{-i\frac{l_A}{2(l_A+l_B)}(y+n\sqrt{\pi})s_B}e^{i(-s_B\hat{p}_B+t_B\hat{q}_B)}|\overline{+}\rangle_B\Bigg)
\end{align}

\section{Shifted error wavefunction \label{logical_error_in_shifted_efn}}

Consider the example of an approximate GKP qubit state whose underlying GKP qubit state is the $|\overline{+}\rangle$, and error wavefunction has non-zero mean displacements given by $u'$ and $v'$, i.e., 
\begin{align}
|\widetilde{\psi}\rangle&=\int du dv \frac{e^{-\frac{(u-u')^2}{2\delta^2}-\frac{(v-v')^2}{2\kappa^2}}}{\sqrt{\pi \kappa\delta}}e^{(i(-u \hat{p} + v \hat{q}))}|\overline{+}\rangle,\\
&=\int du dv \frac{e^{-\frac{(u-u')^2}{2\delta^2}-\frac{(v-v')^2}{2\kappa^2}}}{\sqrt{\pi \kappa\delta}}e^{\frac{iuv}{2}} \hat{X}(u)\hat{Z}(v)|\overline{+}\rangle\\
&=\sum_{n=-\infty}^\infty\frac{1}{\sqrt{\pi \kappa\delta}}\int dv\int du e^{\frac{iuv}{2}} e^{-\frac{(u-u')^2}{2\delta^2}}e^{-iu(2n\sqrt{\pi}+v)}\nonumber\\
&\times  \exp(-\frac{(v-v')^2}{2\kappa^2})\hat{Z}(2n\sqrt{\pi}+v)|p=0\rangle
\end{align}
We have
\begin{align}
|\widetilde{\psi}\rangle&=\sum_{n=-\infty}^\infty\frac{\sqrt{2\pi\delta^2}}{\sqrt{\pi \kappa\delta}}\int dv\int du \frac{e^{-\frac{(u-u')^2}{2\delta^2}}}{\sqrt{2\pi\delta^2}}\nonumber\\
&\times e^{-iu(2n\sqrt{\pi}+v/2)} e^{-\frac{(v-v')^2}{2\kappa^2}}\hat{Z}(2n\sqrt{\pi}+v)|p=0\rangle\\
&=\sqrt{\frac{2\delta}{\kappa}}\sum_n\int dv e^{-\frac{(2n\sqrt{\pi}+v/2)^2}{2/\delta^2}}e^{-iu'(2n\sqrt{\pi}+v/2)}\nonumber\\
&\times e^{-\frac{(v-v')^2}{2\kappa^2}}\hat{Z}(2n\sqrt{\pi}+v)|p=0\rangle\\
&=\sqrt{\frac{2\delta}{\kappa}}\sum_n e^{-\frac{(2n\sqrt{\pi}+v'/2)^2}{2\left(\frac{4+\delta^2\kappa^2}{4\delta^2}\right)}}\int dv  e^{-\frac{\left(v+\frac{2\delta^2\kappa^22n\sqrt{\pi}-4v'}{4+\delta^2\kappa^2}\right)^2}{2\left(\frac{4\kappa^2}{4+\delta^2\kappa^2}\right)}}\nonumber\\
&\times e^{-iu'(2n\sqrt{\pi}+v/2)}\hat{Z}(2n\sqrt{\pi}+v)|p=0\rangle\\
&=\sqrt{2\pi}\sum_n \frac{e^{-\frac{(2n\sqrt{\pi}+v'/2)^2}{2\left(\frac{4+\delta^2\kappa^2}{4\delta^2}\right)}}}{\left(\pi\left(\frac{4+\delta^2\kappa^2}{4\delta^2}\right)\right)^{1/4}}\int dv  \frac{e^{-\frac{\left(v+\frac{2\delta^2\kappa^22n\sqrt{\pi}-4v'}{4+\delta^2\kappa^2}\right)^2}{2\left(\frac{4\kappa^2}{4+\delta^2\kappa^2}\right)}}}{\left(\pi\left(\frac{4\kappa^2}{4+\delta^2\kappa^2}\right)\right)^{1/4}}\nonumber\\
&\times e^{-iu'(2n\sqrt{\pi}+v/2)}\hat{Z}(2n\sqrt{\pi}+v)|p=0\rangle.
\end{align}
In the limit $\kappa^2\delta^2\rightarrow 0$, we get
\begin{align}
|\widetilde{\psi}\rangle&=\sqrt{2\pi}\sum_n \frac{e^{-\frac{\delta^2(2n\sqrt{\pi}+v'/2)^2}{2}}}{(\pi/\delta^2)^{1/4}}\int dv \frac{e^{-\frac{(v-v')^2}{2\kappa^2}}}{(\pi\kappa^2)^{1/4}}\nonumber\\
&\times e^{-iu'(2n\sqrt{\pi}+v/2)}\hat{Z}(2n\sqrt{\pi}+v)|p=0\rangle.
\end{align}
The $p-$quadrature wavefunction of the above state $|\widetilde{\psi}\rangle$ is given by
\begin{align}
    \langle p|\widetilde{\psi}\rangle &=\sqrt{2\pi}\sum_n \frac{e^{-\frac{\delta^2(2n\sqrt{\pi}+v'/2)^2}{2}}}{(\pi/\delta^2)^{1/4}}\int dv \frac{e^{-\frac{(v-v')^2}{2\kappa^2}}}{(\pi\kappa^2)^{1/4}}\nonumber\\
    &\times e^{-iu'(2n\sqrt{\pi}+v/2)}\delta(v-(p-2n\sqrt{\pi}))\\
    &=\sqrt{2\pi}\sum_n \frac{e^{-\frac{\delta^2(2n\sqrt{\pi}+v'/2)^2}{2}}}{(\pi/\delta^2)^{1/4}} \frac{e^{-\frac{(p-2n\sqrt{\pi}-v')^2}{2\kappa^2}}}{(\pi\kappa^2)^{1/4}}\nonumber\\
    &\times e^{-i\frac{u'}{2}(p+2n\sqrt{\pi})}.
\end{align}

When $v'=\sqrt{\pi}$, the state has it's support flipped to that of the ideal GKP qubit state $|\overline{-}\rangle$, however, with an envelope that is still only $\sqrt{\pi}/2$ shifted from the center of the envelope corresponding to the initial $|\widetilde{+}\rangle$ state. Thus, it is not quite $|\widetilde{-}\rangle$. The $p-$quadrature wavefunction description of the state is shown in Fig.~\ref{fig:logical_error_ewfn}.

\bibliography{gkpgraph_references.bib}
\bibliographystyle{unsrt}
\end{document}